\documentclass[10pt]{article}
\usepackage{amsmath,amsthm,amsfonts,amscd,eucal,latexsym,amssymb,bm,amsbsy,mathrsfs} 
\usepackage{epsfig}  
\def\sp{\hskip -5pt} 
\def\spa{\hskip -3pt}

\def\cD{{\ca D}}

\def\cF{{\ca F}}

\def\cH{{\ca H}}

\def\cL{\mathscr{L}}

\def\cS{{\ca S}}

\def\cW{{\ca W}}

\def\sS{{\mathsf S}}

\def\bk{{\bf k}}
\def\bx{{\bf x}}

\def\bC{{\mathbb C}}           

\def\bN{{\mathbb N}}

\def\bR{{\mathbb R}}

\def\bS{{\mathbb S}}

\newsymbol\rest 1316         

\def\gg{{\mathfrak g}} 
\def\gH{{\mathfrak H}}

\def\beq{\begin{eqnarray}}
\def\eeq{\end{eqnarray}}
\def\pa{\partial}
\def\at{\left(}               

\def\ct{\right)}              
\newcommand{\ca}[1]{{\cal #1}}         

\def\ga{\gamma}

\def\la{\lambda}

\def\ph{\varphi}

\def\scri{{\Im^{+}}}
\def\scrim{{\Im^{-}}}

\def\hM{{\widehat{M}}}
\def\hg{{\widehat{g}}}

\newcommand{\nref}[1]{(\ref{#1})}

\newcounter{proposition}[section]
\newcounter{theorem}[section]
\newcounter{lemma}[section]
\newcounter{definition}[section]
\newcounter{corollary}[section]
\newcounter{remark}[section]
\def\theproposition{\thesection.\arabic{proposition}}
\def\thetheorem{\thesection.\arabic{theorem}}
\def\thelemma{\thesection.\arabic{lemma}}
\def\thedefinition{\thesection.\arabic{definition}}
\def\thecorollary{\thesection.\arabic{corollary}}
\def\theremark{\thesection.\arabic{remark}}

\newcommand{\se}[1]{\section{#1}}

\def\vsp{\vspace{0.2cm}}
\def\vspp{\vspace{0.1cm}}

\def\sse #1 {\vsp\ifhmode{\par}\fi\refstepcounter{subsection}
  \noindent {\bf\thesubsection}. {\em #1}.\quad
  \addcontentsline{toc}{subsection}{\protect\numberline{\thesubsection} #1}%
  }

\def\ssb #1 {\vsp\ifhmode{\par}\fi\refstepcounter{subsection}
  \noindent {\bf\thesubsection.} {\em #1.}\quad
  \addcontentsline{toc}{subsection}{\protect\numberline{\thesubsection} #1}%
  }

\def\ssa #1 {\ifhmode{\par}\fi\refstepcounter{subsection}
  \noindent {\bf\thesubsection.} {\bf #1.}\quad
  \addcontentsline{toc}{subsection}{\protect\numberline{\thesubsection} #1}%
  }

\def\proposizione #1 {\vsp\ifhmode{\par}\fi\refstepcounter{proposition}
  \vsp\ifhmode{\par}\fi\noindent {\bf Proposition \theproposition}. \quad {\em #1}}
\def\teorema #1 {\vsp\ifhmode{\par}\fi\refstepcounter{theorem}
  \vsp\ifhmode{\par}\fi\noindent {\bf Theorem \thetheorem}. \quad {\em #1}}
\def\lemma #1 {\vsp\ifhmode{\par}\fi\refstepcounter{lemma}
  \vsp\ifhmode{\par}\fi\noindent {\bf Lemma \thelemma}. \quad {\em #1}}
\def\definizione #1 {\ifhmode{\par}\fi\refstepcounter{definition}
  \vsp\ifhmode{\par}\fi\noindent {\bf Definition \thedefinition}. \quad {\em #1}}
\def\corollario #1 {\vsp\ifhmode{\par}\fi\refstepcounter{corollary}
  \vsp\ifhmode{\par}\fi\noindent {\bf Corollary \thecorollary}. \quad {\em #1}}
  \def\remark {\vsp\ifhmode{\par}\fi\refstepcounter{remark}
  \vsp\ifhmode{\par}\fi\noindent {\bf Remark \theremark}. }

\def\proof #1 {\vspp\ifhmode{\par}\fi\noindent {\it Proof.} {#1} $\Box$\vsp\par}

\begin{document} 
 
\hfill{\sl Desy 07-218 - UTM 718 - ZMP-HH/07-12,  December 2007} 
\par 
\bigskip 
\par 
\rm 
 
 
\par 
\bigskip 
\LARGE 
\noindent 
{\bf  Cosmological horizons and reconstruction of quantum field theories.} 
\bigskip 
\par 
\rm 
\normalsize 
 

\large
\noindent {\bf Claudio Dappiaggi$^{1,a}$},
{\bf Valter Moretti$^{2,3,b}$}, {\bf Nicola Pinamonti$^{1,c}$} \\
\par
\small
\noindent $^1$ 
II. Institut f\"ur Theoretische Physik, Universit\"at Hamburg,
Luruper Chaussee 149, 
D-22761 Hamburg, Germany.\smallskip

\noindent$^2$ Dipartimento di Matematica, Universit\`a di Trento
 and  Istituto Nazionale di Fisica Nucleare -- Gruppo Collegato di Trento, via Sommarive 14  
I-38050 Povo (TN), Italy. \smallskip

\noindent$^3$  Istituto Nazionale di Alta Matematica ``F.Severi''-- GNFM \bigskip

\noindent E-mail: $^a$claudio.dappiaggi@desy.de,
 $^b$moretti@science.unitn.it,  $^c$nicola.pinamonti@desy.de\\ 
 \normalsize


\par 
 
\rm\normalsize 

\rm\normalsize 
 
 
\par 
\bigskip 

\noindent
{\it Dedicated to Professor Klaus Fredenhagen on the occasion of his 60th birthday.}

\bigskip
\noindent 
\small 
{\bf Abstract}. 
As a starting point, we state some relevant geometrical properties enjoyed by
the cosmological horizon of a certain class of 
Friedmann-Robertson-Walker backgrounds. 
Those properties are generalised to a larger class of expanding spacetimes $M$ 
admitting a geodesically complete cosmological horizon $\scrim$ common to all co-moving observers. This 
structure is later exploited in order to recast, in a cosmological background, some recent results for a 
linear scalar quantum field theory in spacetimes asymptotically flat at null infinity. Under suitable
hypotheses on $M$, encompassing both the cosmological de Sitter background and a large class of other FRW 
spacetimes, the algebra of observables for a Klein-Gordon field is mapped into a subalgebra of the 
algebra of observables $\cW(\scrim)$ constructed on the cosmological horizon. There is exactly one pure 
quasifree state $\lambda$ on $\cW(\scrim)$ which fulfils a suitable energy-positivity condition with respect 
to a generator related with the cosmological time displacements. Furthermore $\lambda$ induces a preferred 
physically meaningful quantum state $\lambda_M$ for the quantum theory in the bulk. If $M$  admits a timelike
Killing generator preserving $\scrim$, then the associated self-adjoint generator in the GNS representation 
of $\lambda_M$ has positive spectrum (i.e. energy). Moreover $\lambda_M$  turns out to be invariant under 
every symmetry of the bulk metric which preserves the cosmological horizon. In the case of an expanding 
de Sitter spacetime, $\lambda_M$ coincides with the Euclidean (Bunch-Davies) vacuum state, hence being Hadamard 
in this case. Remarks on the validity of the Hadamard property for $\lambda_M$ in more general spacetimes 
are presented.
\normalsize
\bigskip

\tableofcontents

\se{Introduction}
In the framework of quantum field theory over curved backgrounds we witnessed, in the past few year, an
increased display of new and important formal results. In many cases we can track their origin in the 
existence of a non trivial interplay between some field theories living on a Lorentzian background - say $M$
- and a suitable counterpart constructed over a co-dimension one submanifold of $M$, often chosen as the 
conformal boundary of the spacetime. Usually thought of as a realization of the so-called holographic 
principle, this research line provided its most remarkable results in the framework of (asymptotically) AdS 
backgrounds. As a matter of fact, concepts such as Maldacena's conjecture \cite{Aharony} - in a string 
framework - or Rehren's duality (see \cite{Duetsch} and references therein) - in the algebraic quantum field 
theory setting -  are appearing nowadays almost ubiquitously in the theoretical high-energy physics 
literature. More recently a similar philosophy has been also adopted to deal with a rather different 
scenario, namely asymptotically flat spacetimes, where it is future null infinity -- $\Im^+\sim\bR\times
\bS^2$, {\it i.e.} the conformal boundary -- which plays the role of the above-mentioned co-dimension one 
submanifold \cite{DMP,M1,M2,DA07}. 

Although one could safely claim that all these mentioned results are compelling, one should also actively 
seek connections to those theoretical models which are nowdays testable and, within this respect, one can
safely claim that cosmology is a rather natural playground. In this realm, one of the most widely known 
theories is inflation where, as in other models, the pivotal role is played by a single scalar field living 
on  an (almost) de Sitter background. 
Although, within this framework, most of the results are mainly, though not only, at a classical level, 
it is to a certain extent mandatory to look for a deep-rooted analysis of the full-fledged underlying 
quantum field theory in order to achieve a more firm understanding of the model under analysis. 

To this avail, the first, but to a certain extent, not appealing chance is to perform a case-by-case analysis
of the quantum structure of all the possible models nowadays available. In our opinion a more attractive 
possibility is to look for some mean allowing us to draw some general conclusions or to point out some
universal feature, independently from the chosen model or from the chosen background. Taking into account 
this philosophy, a natural ``first step" to undertake would be to try to implement the previously discussed
bulk-to-boundary correspondence which appears to encode, almost per construction, all the criteria of
universality we are seeking for, in the case of a large class of cosmological models. 

As a starting point point let us assume the {\it Cosmological Principle} which leads the underlying 
background to be endowed with the widely-used Friedmann-Robertson-Walker (FRW) metrics.
A direct inspection of the geometric properties of these spacetimes points out that, in most of the relevant
physical cases, such as de Sitter to quote just one example, it exists a natural submanifold which, at first
glance, appears to be a good candidate as the preferred co-dimension $1$
hypersurface: the {\em cosmological (future or past) horizon} as defined by Rindler \cite{Rindler}. 
More precisely, in this paper we shall consider the cosmological past horizon $\scrim$, in common with all the co-moving 
observers, in order to deal with expanding universes. 
The first of the main aims of this manuscript is indeed to discuss some non
trivial geometric features of the cosmological horizon $\scrim$. Particularly, under some technical 
restrictions on the analytic form of the expanding factor in the FRW metric with flat spatial section, the 
horizon has a universal structure and, hence, it represents the natural setting where to stage a 
bulk-to-boundary correspondence. An expanding universe admits a preferred future-oriented 
timelike vector field $X$ defining the worldlines of co-moving observers, whose common expanding rest-frames
are the $3$-surfaces orthogonal to $X$. In FRW metrics $X$ is a conformal Killing field which becomes 
tangent to the cosmological horizon and, in the class of FRW metrics we consider, it individuates complete 
null geodesics on $\scrim$.

This extent will be generalised to expanding spacetimes $M$ equipped with a geodesically complete cosmological horizon  
$\scrim$ and an asymptotical conformal Killing field $X$,
generally different from FRW spacetimes.
The leading role of $X$ in such a construction is strengthened by its intertwining relation with the 
conformal factor which is a primary condition to take into account if one wants to study the
structure of the {\em symmetry group of the horizon} (actually a subgroup of the huge full isometry group of
the horizon viewed as a semi-Riemannian manifold). We also address such an issue and we 
discover that such a group is actually an infinite dimensional group $SG_\scrim$ which has the
structure of an iterated semidirect product {\it i.e.} it is $SO(3)\ltimes\left(C^\infty(\bS^2)\ltimes
C^\infty(\bS^2)\right)$ where $SO(3)$ is the special orthogonal group with a three dimensional algebra,
whereas $C^\infty(\bS^2)$ stands for the set of smooth functions over $\bS^2$ thought as an Abelian group
under addition. The geometric interpretation of $SG_\scrim$ is intertwined to the following result.
The subgroup of isometries of the spacetime which preserves
the cosmological horizon structure is injectively mapped to a subgroup of $SG_\scrim$ which, hence, encodes 
some of the possible symmetries of the spacetime.
 However it must be remarked that $SG_\scrim$ is universal in the sense that it does not depend on 
the particular spacetime $M$ in the class under consideration.


As a result we find that, under suitable hypotheses on $M$ -- 
valid, in particular,  for certain FRW spacetimes which are de Sitter asymptotically -- the algebra of
 observables $\cW(M)$ of a Klein-Gordon field 
in $M$ is one-to-one (isometrically) mapped to a subalgebra of the algebra of observables $\cW(\scrim)$ 
naturally constructed on the cosmological horizon.  In this sense information of quantum theory in the bulk $M$ is 
encoded in the quantum theory defined on the boundary $\scrim$. 
 It turns out that 
 there is exactly one pure quasifree state $\lambda$ on $\cW(\scrim)$
which fulfils a certain energy-positivity condition with respect to some generators of $SG_\scrim$. The relevant generators 
are here those which can be interpreted as limit values on $\scrim$ of timelike Killing vectors of $M$, whenever one fixes a
 spacetime  $M$ admitting $\scrim$ as the cosmological horizon.
However, exactly as the geometric structure of $\scrim$,  $\lambda$ is universal 
in the sense that it does not depend on the particular spacetime $M$ in the class under consideration.
The GNS-Fock representation of $\lambda$ individuates a unitary irreducible representation of 
$SG_\scrim$.
Fixing an expanding spacetime $M$ with complete cosmological horizon,
$\lambda$ induces a  preferred quantum state $\lambda_M$ 
for the quantum theory in $M$ and it enjoys 
remarkable properties. It turns out to be invariant under all those isometries of $M$ (if any) that preserve
the cosmological horizon structure. If $M$ admits a timelike Killing generator preserving $\scrim$, the associated
 self-adjoint generator in the GNS representation of $\lambda_M$ has positive spectrum, {\it i.e.}, energy. 
Eventually, if $M$ is the expanding de Sitter spacetime,
 $\lambda_M$ coincides to the Euclidean (Bunch-Davies) vacuum state, so that it is Hadamard in that case at least.
Actually, Hadamard property seems to be valid in general, but that issue will be investigated elsewhere.

As a final technical remark we would like to report that in the derivation of many  
results reported here we have been guided by  similar analyses previously performed 
in the case of asymptotically flat spacetime, using the null infinity as co-dimension one submanifold.
However, to follow the subsequent discussion there is no need of being familiar with the tricky notion 
of asymptotically flat spacetime.\\
%

\ssa{Notation, mathematical conventions}\label{secgauge} 
Throughout $\bR^+:= [0,+\infty)$, $\bN:= \{0,1,2,\ldots\}$. For  smooth manifolds $M,N$, 
$C^\infty(M;N)$
(omitting $N$ whenever $N=\bR$) is the space of smooth functions $f: M\to N$.
$C^\infty_0(M;N)\subset C^\infty(M;N)$ is the subspace of compactly-supported functions.  
If $\chi : M\to N$ is a diffeomorphism, $\chi^*$ is the natural extension to tensor bundles 
(counter-, co-variant and mixed) from $M$ to $N$ (Appendix C in \cite{Wald}).
A spacetime $(M,g)$ is a  Hausdorff, second-countable, smooth, four-dimensional  
connected manifold $M$, whose smooth metric has signature $-+++$. We shall also assume that 
a spacetime is oriented and time oriented.
We adopt definitions of causal structures of Chap. 8 in \cite{Wald}.
  If $S\subset M\cap \hM$, $(M,g)$ and $(\hM,\hg)$ being spacetimes, 
  $J^\pm(S;M)$ ($I^\pm(S;M)$) and $J^\pm(S;\hM)$
 ($I^\pm(S;\hM)$) indicate the causal (chronological) 
 sets associated to $S$ and respectively referred to the spacetime 
 $M$ or $\hM$. \\

\ssa{Outline of the paper} In section 2 we introduce and discuss the geometric set-up of the backgrounds 
we are going to take into account throughout this paper. Particularly we find under which analytic conditions 
on the expanding factor, a Friedmann-Robertson-Walker (FRW) spacetime can be smoothly extended to a larger 
spacetime that encompasses the cosmological horizon.
%
In section 3 we provide a generalisation of
the results of section 2 and we study their implications. Furthermore we introduce and discuss
the structure of the horizon symmetry group showing its interplay with the possible isometries of the bulk
metric. In section 4 we study the structure of bulk
scalar QFT and of the associated Weyl algebra and its the horizon counterpart. Furthermore we discuss the 
existence of a preferred  algebraic state invariant under the full 
symmetry group, which enjoys some uniqueness/energy-positivity properties.
Subsections 4.3 and
4.4 are devoted to the development of the interplay between the bulk and the boundary theory; a particular
emphasis is given to the selection of a natural preferred bulk states and on the analysis of its properties.
Since all these conclusions are based upon some a priori assumptions on the behaviour of the solutions in the
bulk of the Klein-Gordon equation with a generic coupling to curvature, we shall devote section 4.5 to test 
these requirements. Eventually, in section 5, we draw some conclusions and we provide some hints on future
research perspectives.\\

\se{Cosmological horizons and asymptotically flatness}
\ssb{Friedmann-Robertson-Walker spacetime and cosmological horizons}
A homogeneous and isotropic universe can be locally described by a smooth
spacetime, in the following indicated by $(M,g_{FRW})$, where $M$ is a smooth
Lorentzian manifold equipped with the following {\bf Friedmann-Robertson-Walker} (FRW) {\bf metric}
\beq\label{metric}
g_{FRW} = -dt \otimes dt +a(t)^2\left[ \frac{1}{1- \kappa r^2} dr\otimes dr+r^2d
\bS^2(\theta,\varphi)\right].
\eeq
Above, $d\bS^2(\theta,\varphi)=d\theta\otimes d\theta +\sin^2\theta \:d\phi \otimes d\phi$ is the standard
metric on the unit $2$-sphere and, 
up to normalisation, $\kappa$ can take the values $-1,0,1$ corresponding respectively 
to an hyperbolic, flat and closed spaces. The coordinate $t$ ranges in some open interval $I$. Here $a(t)$ 
is a smooth  function of $t$ with constant sign (since $g$ is nondegenerate).
Henceforth we shall assume that $a(t)>0$ when $t \in I$. We also suppose that the field $\partial_t$ 
individuates the time orientation 
of the spacetime.\\
Physically speaking and in the universe observed nowadays, the sections of $M$ at fixed $t$  
are the isotropic and homogeneous $3$-spaces  
containing the matter of the universe, the world lines describing the histories of those particles of matter being 
integral curves of $\partial_t$. 
In this picture, the {\bf cosmic time} $t$ is the {\em proper-time} measured at rest with each 
of these particles, whereas the scale $a(t)$ measures the size of the observed cosmic expansion in function of $t$.\\
The metric \eqref{metric} may enjoy two physically important features. Consider a 
co-moving observer pictured by a  integral line $\gamma=\gamma(t)$, $t\in I$, of the field $\partial_t$ and 
focus on $J^-(\gamma)$.
If $J^-(\gamma)$ does not cover the whole spacetime $M$, the 
observer $\gamma$ cannot receive physical information from some events of $M$ during his/her history: causal 
future-directed signals starting from  $M \setminus  J^-(\gamma)$ cannot achieve any point on $\gamma$. 
 In other words, and adopting the terminology of \cite{Rindler},
  a {\bf  cosmological horizon} takes place for $\gamma$ and it is the null 
 $3$-hypersurface $\partial  J^-(\gamma)$. Conversely, whenever
$J^+(\gamma)$ does not cover the whole spacetime $M$, physical information sent 
by the observer $\gamma$ during his/her story is prevented from getting to some events of $M$: 
Causal future-directed signals starting from $\gamma$  
do not reach any point in  $M \setminus J^+(\gamma)$.  In this case, exploiting again the terminology of 
\cite{Rindler}, a {\bf cosmological past horizon} exists for $\gamma$. It is the null $3$-hypersurface 
$\partial J^+(\gamma)$.\\
As it is well-known, a sufficient condition for the appearance of cosmological horizons can be obtained from the
following analysis. One re-arranges the metric \eqref{metric} into the form
\beq\label{cosmo0}
g_{FRW}=a^2(\tau)\left[-d\tau\otimes d\tau +\frac{1}{1-\kappa r^2}dr\otimes dr+r^2d
\bS^2(\theta,\varphi)\right]\doteq a^2(\tau)g(\tau,r,\theta,\varphi),
\eeq
where \beq \label{tau}\tau(t)=  d + \int a^{-1}(t)dt
\eeq is the {\bf conformal cosmological time},
 $d \in \bR$ being any fixed constant. By construction  $\tau=\tau(t)$ is a 
diffeomorphism from $I$ to some open, possibly infinite, interval $(\alpha,\beta)\ni \tau$. 
Notice both that $\partial_\tau$ is a conformal Killing vector field whose integral lines
coincide, up to the parametrisation, to the integral lines of $\partial_t$ and that $(M,g_{FRW})$ is globally 
hyperbolic.

As causal structures are preserved under conformal rescaling of the metric, a straightforward analysis based on 
the shape of $g$ in \eqref{cosmo0} establishes that $J^-(\gamma)$ does not cover
he whole spacetime $M$ whenever $\beta<+\infty$. In that case a  cosmological event horizon takes place for 
$\gamma$. Similarly $J^+(\gamma)$ does not cover the whole spacetime $M$ whenever
$\alpha>-\infty$. In that case a  cosmological past horizon takes place for $\gamma$.
 In both cases the horizons $\partial J^-(\gamma)$ and $\partial J^+(\gamma)$ are null $3$-hypersurfaces 
diffeomorphic to $\bR \times \bS^2$, made of null geodesics of $g_{FRW}$. One may think of these surfaces as
the limit light-cones emanating from $\gamma(t)$,
 respectively towards the past or towards the future,  as $t$ tends to $\sup I$ or 
 $\inf I$ respectively. The tips of the cones 
 generally get lost in the limit procedure: In realistic models $\alpha$ and $\beta$ correspond, when they 
 are finite, to a {\em big bang} or a {\em big crunch}  respectively. As a general comment, we stress that 
 {\em the cosmological horizons introduced above generally  depend on the fixed observer $\gamma$}.
 \remark \label{remcentr}  The requirement on the finiteness of the bounds $\alpha$ and $\beta$ for the range 
 of the conformal cosmological time $\tau$ are {\em sufficient} conditions for the existence of the 
 cosmological horizons, but they are by no means necessary. Indeed it may happen  that -- and this is the
 case of  de Sitter spacetime -- there is, indeed a cosmological horizon {\em arbitrarily close to $M$, but 
 outside $M$}. This happens when the spacetime $M$ and its metric can be extended beyond its original region 
 $M$ to a larger spacetime $(\hM,\hg)$ so that it happens that
$\scri = \partial J^-(M; \widehat{M}) = \partial M$ and $\scrim = \partial J^+(M; \widehat{M})= \partial M$.
Hence the cosmological horizon $\Im^+$ or $\Im^-$ coincides with the boundary $\partial M$ and, by 
construction, it  does not depend on the considered observer $\gamma$ (an integral curve of the field 
$\partial_t$) evolving in $M$. Referring in particular to a conformally static region $M$ (equipped with the 
metric (\ref{metric}) for $\kappa =0$) embedded in the complete  de Sitter spacetime $\hM$, $\partial M$ 
turns out to be a null surface with the topology of  $\bR\times \bS^2$. In the following we shall focus on 
this type of cosmological horizons.

\ssb{FRW metrics with $\kappa=0$ and associated geometric structure}\label{FRWsection}
Here, we would like to pinpoint some geometrical properties enjoyed by a subclass if the FRW spacetimes
that will be used later in order to get the main results presented in this paper.
To this end we consider here the spacetime $(M,g_{FRW})$, where $M\simeq(\alpha,\beta)\times\bR^3\:$ and the 
metric $g_{FRW}$ is like in (\ref{cosmo0}), but with $\kappa=0$.

\begin{figure}[!t]
\centering
\includegraphics[bb= 0 0 202 202, scale=.8]{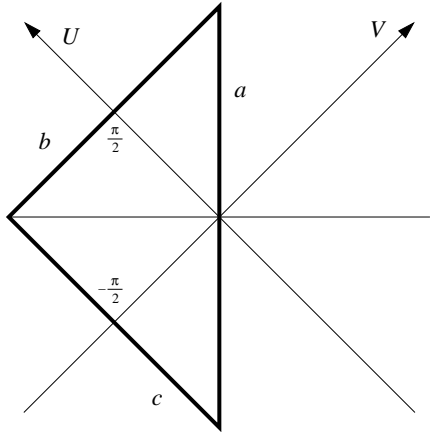}
\caption{The interior of the triangle represents the original FRW background seen as an open subset of
Einstein's static universe. Each point in the $(U,V)$-plane represents a 2-sphere and, furthermore, the 
segments $b$ and $c$ are respectively $\scri$ and $\scrim$.}
  \label{DMP2figura}
\end{figure}

\noindent
Furthermore we shall restrict our attention to the case where the factor $a(\tau)$ in \nref{cosmo0} has the following form
\beq \label{condag}
a(\tau) = \frac{\gamma}{\tau} + O\at \frac{1}{\tau^2} \ct\:, \quad
\frac{d a(\tau)}{d \tau} = -\frac{\gamma}{\tau^2} + O\at \frac{1}{\tau^3} \ct\:
\eeq
for either $(\alpha,\beta):= (-\infty,0)$ and $\gamma<0$, 
or $(\alpha,\beta):= (0,+\infty)$ and $\gamma>0$. The above asymptotic values are meant to be taken as $\tau
\to -\infty$ or $\tau \to +\infty$ respectively.
The first issue we are going to discuss is the extension of the spacetime $(M, g_{FRW})$ to a larger
spacetime $(\hM,\hg)$ that encompasses $\scri$ and/or $\scrim$. To this end, if we introduce the new coordinates $U=\tan^{-1}(\tau+r)$ and 
$V=\tan^{-1}(\tau-r)$ ranging in subsets of $\bR$ individuated by $\tau \in (\alpha,\beta)$ and $r\in (0,+
\infty)$, \eqref{cosmo0} can be written as:
\beq\label{metcomp}
g_{FRW}=\frac{a^2(\tau(U,V))}{\cos^2 U\cos^2 V}\left[-\frac{1}{2}dU\otimes dV - \frac{1}{2}dV \otimes dU+
\frac{\sin^2(U-V)}{4}d\bS^2(\theta,\varphi)\right].
\eeq
The metric, obtained cancelling the overall factor 
$a^2(\tau(U,V))/(\cos^2 U\cos^2 V)$, is well-behaved and  smooth for  $U,V \in \bR$ removing the axis $U=V$. 
This is nothing but the apparent singularity appearing for $r=0$ in the original metric 
(\ref{cosmo0}). Consider $\bR^2$ equipped with null coordinates
$U,V$ with respect to the standard Minkowskian metric on $\bR^2$ and assume that every point is a $2$-sphere 
with radius $|\sin(U-V)|/2$ (hence the spheres for $U=V$ are degenerate). Then, let us focus on the segments 
in $\bR^2$
$$\begin{array}{l}
a,\; V=U\; \textrm{with}\; U\in (-\pi/2,\pi/2),\\
b,\; U=\pi/2\;  \textrm{with}\; V\in (-\pi/2,\pi/2),\\
c, V=-\pi/2\;  \textrm{with}\;U\in (-\pi/2,\pi/2).
\end{array}$$
\noindent The original spacetime $M$ is realized as a suitable subset of the union of the segment $a$, 
{\it i.e.} $r=0$, and the {\em interior} of the triangle $abc$, {\it i.e.} $r>0$, as in the figure 
\ref{DMP2figura}. 
In this picture it is natural to assume that the null {\em endless} segments $b$ and $c$ representing null 
$3$-hypersurfaces diffeomorphic to $\bR\times\bS^2$, individuate respectively $\scri$ and $\scrim$ provided 
that $\beta=+\infty$ in the first case and/or $\alpha = -\infty$ in the second case where $(\alpha,\beta)$ is
the domain of $\tau$. Otherwise the points of $M$ cannot get closer and closer to all the points of those 
segments. Therefore we are committed to assume $\alpha=-\infty$ and/or $\beta=+\infty$ and we stick with this 
assumption in the following discussion.\\
%
%
%
%
Summarising, we wish to extend $g_{FRW}$ smoothly to a region larger than the open triangle $abc$ 
joined with $a$, and including  one of the endless segments $b$ and $c$ at least.
In the case $a(\tau)$ is of the form \nref{condag}, the function $a^2(\tau(U,V))/(\cos^2 U\cos^2 V)$ 
is smooth in neighbourhoods of the open segments $b$ and $c$ only if $\ga\neq 0$, and in particular
it does not vanish on $b$ and $c$, making nondegenerate $\hg$ thereon. 
However, a bad singularity appears as soon as $U=-V$, that is $\tau =0$. Therefore either:

 $(\alpha,\beta) = (0,+\infty)$ -- and in this case $M$ ($r\geq 0$, $\tau \in (0,+\infty)$) 
coincides with the upper half of the triangle $abc$, and it may be extended to a larger spacetime $(\hM, \hg)$ 
by adding a neighbourhood of the endless segment $b$  viewed as $\scri$  -- or
 
 $(\alpha,\beta) = (-\infty,0)$ -- and in this case $M$ ($r\geq 0$, $\tau \in (-\infty,0)$) 
coincides to the lower half of the triangle $abc$,
and it may be extended to a larger spacetime $(\hM, \hg)$ by adding a neighbourhood the endless segment $c$ 
viewed as $\scrim$.\\
\noindent In both cases the line $U=-V$ does not belong to $M$ and to its extension, and the metric $\hg$
coincides with the right-hand side of \eqref{metcomp}.

The function $a(\tau)$ and its interplay with the vector field $\pa_\tau$ when approaching the cosmological 
horizon will play a distinguished role in our construction for this reason let's enumerate below some of its
properties that we are going to generalise in the next section.
To this end, notice that $a(\tau)$ is smooth in $\hM$ and vanishes exactly either on 
$\Im^+=\partial J^-(M; \widehat{M})$ or on $\Im^-=\partial J^+(M; \widehat{M})$, depending on the
considered values for the interval $(\alpha,\beta)$ and for $\gamma$ as discussed below formula 
\eqref{condag}. On the other hand, by direct inspection
\beq
da\rest_{\scri} = -2\gamma dU\: \label{domegaU},\quad
da\rest_{\scrim} = -2\gamma dV,\:.\label{domegaV}
\eeq
and hence $da$ does not vanish either on $\Im^+$ or on $\Im^-$, provided $\ga \neq 0$.
By direct inspection one finds that,  restricting either to $\scri$ or $\scrim$, the metric $\hg$ takes the 
following distinguished form called {\bf Bondi form}:
\beq 
\widehat{g} \rest_{\Im^\pm}   = \gamma^2 \left(- d\ell \otimes  d a -   d a \otimes d\ell + d\bS^2(\theta,
\varphi)\right)
\nonumber\:,
\eeq
where, with $\Im^\pm$, it is implicitly assumed that one must choose either $\Im^+$ or $\Im^-$ and where, for
arbitrarily fixed constants $k_+,k_-$ 
$$
\ell(U) =  - \gamma^{-1} \tan U   + k_-\quad\text{ on } \scrim\; , \qquad \ell(V) = - \gamma^{-1} \tan V   + 
k_+ \quad\text{ on  }\scri \;,
$$
hence $\ell \in \bR$ turns out to be the parameter of the integral lines of $n\doteq\nabla a$.\\
Consider then the vector field $\pa_\tau$, it is an easy task to check that it is a 
conformal Killing vector for $\widehat{g}$ in $M$  with conformal Killing equation  
$$
\cL_{\partial_\tau} \hg = -2 \partial_\tau(\ln a)  \: \hg\:.
$$
where the right-hand side vanishes approaching either $\Im^+$ or $\Im^-$.
Furthermore, $\partial_\tau$ tends to become tangent to either 
$\Im^+$ or $\Im^-$ approaching it and it coincides to $-\gamma \widehat{\nabla}^b a$ thereon, as can be 
directly seen from the form of $\ell$.

\section{Expanding universes with cosmological horizon and its group.}

\ssb{Expanding universes with cosmological horizon $\scrim$}
The previous discussion remarked that in an expanding FRW spacetimes the scale factor $a$ 
and its interplay with the conformal Killing field $\partial_\tau$ play a distinguished role
when approaching the cosmological horizon.
A reader interested in asymptotically flat spacetime could have noticed that many of the above mentioned geometrical properties
are shared by the structure of null infinity.
In that realm, in \cite{DMP,M1,M2}, it was shown that, when dealing with quantum field theory issues, a key role is played by a 
certain symmetry group of diffeomorphisms defined on $\scri$, the so called BMS group, which has the most 
notable property to embody the isometries of the bulk spacetime \cite{Geroch,AX} through a suitable geometric
correspondence of generators.  
In the following we first generalise the result presented in the section \ref{FRWsection}
and then we shall construct the counterpart of 
the BMS group for the found class of spacetimes and the particular form of cosmological horizons.
\definizione \label{defexp}
{\em A globally hyperbolic spacetime $(M,g)$ equipped with a positive smooth function 
$\Omega: M \to \bR^+$,  a future-oriented timelike  vector $X$ defined on $M$, and a constant $\gamma\neq 0$,
will be called an {\bf expanding universe with (geodesically complete) cosmological (past) horizon} when the 
following facts hold:

\begin{enumerate}

\item {\bf Existence and causal properties of horizon}. $(M,g)$ can be isometrically embedded as the interior 
of a sub manifold-with-boundary of a larger spacetime $(\widehat{M}, \hg)$, 
the boundary $\scrim := \partial M$  verifying $\scrim \cap J^+(M; \widehat{M}) = \emptyset$.

\item {\bf Data interplay 1)}. $\Omega$ extends to a smooth function on $\widehat{M}$ such that (i)
$\Omega\spa \rest_\scrim =0$ and (ii) $d\Omega  \neq 0$ everywhere on $\Im^-$.

\item {\bf Data interplay 2)}. $X$ is a conformal Killing vector for $\hg$ in a neighbourhood of $\scrim$ 
in $M$, with
\beq \cL_X(\hg)= - 2  X(\ln \Omega)\: \hg\:, \label{confscri}\eeq
where (i) $X(\ln \Omega) \to 0$ approaching $\scrim$  and (ii) $X$ does not tend everywhere to the zero vector
approaching $\scrim$ .

\item {\bf Global Bondi-form of the metric on $\scrim$ and geodesic completeness}.   (i) $\scrim$ is diffeomorphic to $\bR \times \bS^2$, 
(ii) the metric $\hg\rest_\scrim$ takes the {\bf Bondi form} globally up to the constant factor
$\gamma^2>0$:
\beq
\hg\spa \rest_\scrim = \gamma^2 \left(-d\ell \otimes d\Omega  -d\Omega \otimes d\ell + d \bS^2(\theta,\phi)\right)\:,
\quad \ell\in \bR\:,\: (\theta,\phi)\in \bS^2 \:,\:\Omega =0\label{quasih}
\eeq
$d\bS^2$ being the standard metric on the unit $2$-sphere. Hence $\scrim$ is a null $3$-submanifold, and 
(iii) the  curves $\bR \ni \ell \mapsto (\ell, \theta,\phi)$ are complete null $\hg$-geodesics.

\end{enumerate}
The manifold $\scrim$ is called the {\bf cosmological (past) horizon of $M$}. The integral parameter of $X$
is called the {\bf conformal cosmological time}.
 There is a completely analogous definition of {\bf contracting universe}
referring to the existence of  $\scri$ in the future instead of $\Im^-$.}

\remark \label{remarkgeo}\\ 
{\bf (1)} In view of condition 3, the vector $X$ is a Killing vector of the metric $g_0:= \Omega^{-2} g$ in a 
neighbourhood of $\scrim$ in $M$. In such a neighbourhood, one can think of $\Omega^2$ as an expansion scale 
evolving with rate $X(\Omega^2)$ referred to the conformal cosmological time. \\
{\bf (2)} $\scrim \cap J^+(M; \widehat{M}) = \emptyset$ entails $M= I^+(M;\hM)$ 
and $\scrim=\partial M=\partial I^+(M;\hM)=\partial J^+(M;\hM)$, so that $\scrim$ has the proper 
interpretation as a past cosmological horizon in common for all the observers in $(M,g)$ evolving along 
the integral lines of $X$.\\
{\bf (3)} It is worth stressing that the spacetimes considered in the given definition are neither homogeneous nor
isotropic in general; hence we can deal with a larger class of manifolds than simply the FRW spacetimes.\\

\noindent Similarly to the particular case examined previously, also in the general case pictured by
Definition \ref{defexp}, the conformal Killing vector field $X$ becomes tangent to $\scrim$ and it coincides with $\partial_\ell$
up to a nonnegative factor, which now may depend on angular variables, as we go to establish. The proof of
the following proposition is in the Appendix.

\proposizione \label{X} {\em If $(M,g, \Omega, X,\gamma)$ is an expanding universe with cosmological horizon,
the following holds. 

{\bf (a)} $X$  extends smoothly to a unique smooth vector field $\widetilde{X}$ on $\scrim$,  which may 
vanish on a closed subset of $\scrim$ with empty interior at most. Then $X$ fulfils the $\hg$-Killing 
equation on $\scrim$. 

{\bf (b)} $\widetilde{X}$ has the form $f\partial_\ell$, where,
 referring to the representation $\scrim \equiv \bR \times \bS^2$,  $f$ depends only on the variables  
 $\bS^2$  and, furthermore, it is smooth and nonnegative.}\\

\noindent Since, for the FRW spacetimes, the function $f=f(\theta,\phi)$ appearing in $\widetilde{X}= f\partial_\ell$
takes the constant value $1$, the presence of a nontrivial function $f$ is related to the failure of isotropy for the more general spacetimes considered in Definition \ref{defexp}.\\

\ssb{The horizon symmetry group $SG_\scrim$}
In the forthcoming discussion we shall make use  several times of the following technical fact. In
the representation $\scrim \equiv \bR \times \bS^2 \ni (\ell, s)$,
 the null $\hg$-geodesic segments imbedded in $\scrim$ are all of  the curves
\beq J \ni  \ell \mapsto (\alpha \ell+ \beta, s)\:, \quad \mbox{for constants $\alpha\neq 0, \beta\in \bR$, $s \in \bS^2$, 
and some interval $J\subset
\bR$.} \label{geod}\eeq
In this section, in the hypotheses of definition \ref{defexp}, we select a subgroup $SG_\scrim$ of physically relevant 
isometries of $\scrim$. We shall see in Proposition \ref{Gscrim} that, as matter 
of fact, $SG_\scrim$ contains the isometries generated by Killing vectors obtained as a limit towards 
$\scrim$ of (all possible) Killing vectors of $(M,g)$,  when these vectors tend to become tangent to 
$\scrim$. As a preliminary proposition, it holds:

\proposizione \label{togroup} {\em If $(M,g, \Omega, X, \gamma)$ is an expanding universe with cosmological 
horizon and $Y$ is a Killing vector field of $(M,g)$, $Y$ can be extended to a smooth vector field 
$\widehat{Y}$ defined on $\hM$ and 

{\bf (a)} $\cL_{\widehat{Y}}\widehat{g} =0$ on $M \cup \scrim$;

{\bf (b)} $\widetilde{Y}:= \widehat{Y}\spa\rest_\scrim$ is uniquely determined by $Y$, and it is tangent to 
$\scrim$ if and only if $g(Y,X)$ vanishes approaching $\scrim$ from $M$. Restricting to the linear space of 
the Killing fields $Y$ on $(M,g)$ such that $g(Y,X)\to 0$ approaching $\scrim$, the following further facts
hold.

{\bf (c)} If $\widetilde{Y}$  vanishes in some $A \subset \scrim$ and $A \neq \emptyset$ is open 
with respect to the topology of $\scrim$, then $Y=0$ everywhere in $M$ as well as $\widehat{Y}$ in $M\cup 
\scrim$.  

{\bf (d)} The linear map $Y \mapsto \widetilde{Y}$ is injective, {\it i.e.} Killing vectors of $(M,g)$ are 
represented on $\scrim$ {\em faithfully}\\}

\noindent The proof of the proposition  above is given in the Appendix.\\
The statements (a) and (b) of Proposition \ref{togroup} establish that the Killing vectors $Y$ in $M$ with 
$g(Y,X) \to 0$ approaching $\scrim$
extend to {\em Killing
vectors} of $(\scrim,h)$, $h$ being the degenerate metric on $\scrim$ induced by
$\hg$. 

Since the vector fields $\widehat{Y}$ tangent to $\scrim$ admit $\scrim$ as invariant manifold, we can define

\definizione \label{defpreservscri}
{\em If $(M,g, \Omega, X, \gamma)$ is an expanding universe with cosmological horizon,  a Killing vector field of $(M,g)$, $Y$, is said to 
{\bf to preserve $\scrim$} if  $g(Y,X)\to 0$ approaching $\scrim$. Similarly, the Killing isometries of the (local) one-parameter group generated by 
$Y$ are said {\bf to preserve $\scrim$}.}\\

\noindent In the rest of this part we shall consider the one-parameter group of isometries 
of $(\scrim,h)$ generated by such Killing vectors $\widehat{Y}\spa \rest_\scrim$. These isometries amount to 
a little part of the huge group of isometries of $(\scrim,h)$. For instance, referring to the 
representation $(\ell, s)\in \bR \times \bS^2 \equiv \scrim$, for {\em every} smooth diffeomorphism $f:\bR\to
\bR$, the transformation $\ell\to f(\ell),\quad s\to s\label{sa}$ is an isometry of $(\scrim,h)$. However 
only diffeomorphisms of the form $f(\ell) = a\ell + b$ with $a\neq 0$ can be isometries generated by 
the restriction $\widehat{Y}\spa \rest_\scrim$ to $\scrim$ of extensions of Killing fields $Y$ of $(M,g)$ 
as in the proposition \ref{togroup}.  This is because those isometries 
are restrictions of isometries of the manifolds-with-boundary 
$(M\cup \{\scrim\},\: \hg\spa\rest_{M\cup \{\scrim\}})$, and thus they {\em preserve the null} $\hg$-{geodesics} 
in $\scrim$. These geodesics have the form (\ref{geod}). The requirement that, for every constants 
$a,b\in \bR$, $a\neq 0$, there must be constants $a',b'\in \bR$, $a'\neq 0$ such that $f(a\ell + b) = a'\ell + b'$
for all $\ell$ varying in a fixed nonempty interval $J$, is fulfilled only if $f$ is an affine 
transformation as said above. We relax now the constraints on the above transformations allowing them also to be
dependant on the angular coordinates. Hence we aim to study the class $G_\scrim$ of diffeomorphisms $F:\scrim
\to\scrim$
\beq
\ell \to \ell' := f(\ell, s)\:, \quad s \to s' := g(\ell, s)
\quad \mbox{with $\ell \in \bR$ and $s \in \bS^2$,}\label{diffgen}
\eeq
such that: (i) they are isometries of the degenerate metric $h$ induced by $\hg\rest_{\scrim}$ (\ref{quasih}) 
and (ii) they may be restrictions to $\scrim$ of isometries of $\hg$ in $M\cup \scrim$. \\
Assume that $F\in G_\scrim$. The curve $\gamma: \bR \ni \ell \to \gamma_{s}(\ell) \equiv (\ell, s)$ (with $s\in \bS^2$ 
arbitrarily fixed) is a null geodesic forming $\scrim$, therefore
$\bR \ni \ell \to F (\gamma_s(\ell))$ has to be, first of all, a null curve. In other words
$$\hg\spa\rest_{\scrim}\spa\left(\frac{\partial f}{\partial \ell} \frac{\partial}{\partial \ell}
+ \frac{\partial g}{\partial \ell} \frac{\partial}{\partial \theta} +
\frac{\partial g}{\partial \ell} \frac{\partial }{\partial \phi},
\frac{\partial f}{\partial \ell} \frac{\partial}{\partial \ell}
+ \frac{\partial g}{\partial \ell} \frac{\partial}{\partial \theta} +
\frac{\partial g}{\partial \ell} \frac{\partial }{\partial \phi}\right) =0\:.$$
Using (\ref{quasih}) and arbitrariness of $s\equiv (\theta,\phi)$, it implies that $g$ does not depend on $\ell$ since the standard metric on the unital sphere
is strictly positive definite. The map $g$ has to be an isometry of $\bS^2$ equipped with its standard metric.
In other words $g \in O(3)$. Moreover, $\bR \ni \ell \to F (\gamma_s(\ell)) = (f(\ell, s), g(s))$ 
has to be a null geodesic which belongs 
to $\scrim$. As a consequence of (2) in remark \ref{remarkgeo},
$ (f(\ell, s), g(s)) = (c(s) \ell + b(s), g(s))$ for some fixed numbers $c(s),b(s) \in \bR$ 
with $c(s)>0$, and for every $\ell \in \bR$. 
Summarising, it must be $g(\ell, s) = R(s)$ for all $\ell, s$ and $f(\ell, s)= c(s)\ell + b(s)$, for all 
$\ell, s$, for some $R\in O(3)$, $c,b \in C^\infty(\bS^2)$ with $c(s) \neq 0$. It is obvious that, 
conversely, every such a diffeomorphism fulfils (i) and (ii). 

\remark
{\bf (1)} By direct inspection one sees that the class $G_\scrim$ of all diffeomorphisms $F$ as above is a group with
 respect to the composition of diffeomorphisms.\\ 
{\bf (2)} Only transformations $F\in G_\scrim$, associated with $R$ lying in the component connect to the
identity of $O(3)$, {\it i.e.}, $SO(3)$ belong to a one-parameter group of isometries induced by Killing 
vectors in $M$. 

\noindent From now on we shall restrict ourselves to the subgroup of $G_\scrim$ whose elements are 
constructed using elements of $SO(3)$ and each element of the one-parameter group of diffeomorphisms 
generated by a vector field $Z$ will be denoted by $\exp\{tZ\}$ being $t\in\bR$.

\definizione \label{defSG} {\em The {\bf horizon symmetry group} $SG_{\scrim}$ is the group (with respect to the composition of functions) 
of all diffeomorphisms of $\bR \times \bS^2$,
\beq
F_{(a,b,R)} : \bR \times \bS^2 \ni (\ell, s) \mapsto \left(e^{a(s)}\ell + b(s), R(s)\right) \in \bR \times \bS^2
\quad \mbox{with $\ell \in \bR$ and $s \in \bS^2$,}\label{group}
\eeq
where $a,b \in C^\infty(\bS^2)$ are arbitrary smooth functions and $R \in SO(3)$.\\
The {\bf Horizon Lie algebra} $\gg_{\scrim}$ is 
the infinite-dimensional Lie algebra of smooth vector fields on $\bR \times \bS^2$ generated by the fields 
$$ S_1\:, S_2\:, S_3\:,\: \beta \partial_\ell\:, \:
\ell\alpha \partial_\ell\:, \quad\mbox{for all $\alpha,\beta \in C^\infty(\bS^2)$.}$$
$S_1,S_2,S_3$ indicate the three smooth vector fields on the unit sphere 
$\bS^2$ generating rotations 
about the orthogonal axes, respectively, $x$, $y$ and $z$.}\\

\noindent  It is worth noticing that $SG_{\scrim}$ depends on the geometric structure of $\scrim$ but not on the attached spacetime 
$(M,g)$, which, in principle, {\em could not even admit any Killing vector preserving $\scrim$}. In this 
sense  $SG_{\scrim}$ is a {\em universal} object for the whole class of expanding spacetimes with cosmological horizon.
$SG_{\scrim}$ may be seen as an abstract group defined on the set $SO(3) \times C^\infty(\bS^2)\times 
C^\infty(\bS^2)$, without reference to any expanding spacetime with cosmological horizon $(M,g)$.
Adopting this point of view, if we indicate 
$F_{a,b,R}$ by the abstract triple $(R,a,b)$,  the composition between elements in  $SG_\scrim$ reads
\beq
(R, a,b) (R', a',b') = \left(RR',\:\: a' + a \circ R' ,\:\: e^{a\circ R'} b' + b\circ R' \right),\;
\label{product}  
\eeq
for any $(R,a,b), (R', a',b') \in SO(3) \times C^\infty(\bS^2)\times C^\infty(\bS^2)$ and where $\circ$ 
denotes the usual composition of functions. 

\noindent The relationship between $SG_\scrim$ and $\gg_\scrim$ is clarified in the following proposition.

\proposizione \label{Gscrim} {\em Referring to the definition \ref{defSG},
the following facts hold:

{\bf (a)}  Each vector field $Z \in \gg_\scrim$ is complete and the generated global one-parameter group of diffeomorphisms 
of $\bR \times \bS^2$, $\{\exp\{tZ\}\}_{t\in \bR}$,
is a subgroup of $SG_\scrim$.
 
{\bf (b)} For every $F \in SG_\scrim$ there are  $Z_1,Z_2 \in \gg_\scrim$ -- with, possibly, $Z_1=Z_2$ -- 
such that  $F= \exp\{t_1Z_1\} \exp\{t_2Z_2\}$ for some real numbers $t_1,t_2$.}\\

\noindent The proof of this proposition is in the Appendix.\\
Furthermore, we have the following important result which finally
 makes explicit the interplay between Killing vectors $Y$ in $M$ preserving $\scrim$, the group $SG_{\scrim}$
 and the Lie algebra $\gg_\scrim$.

\teorema \label{theorem2} {\em Let  $(M,g, \Omega, X, \gamma)$ be an expanding universe with cosmological horizon and
 $Y$ a Killing  vector field of $(M,g)$ preserving $\scrim$.
The following  holds.

{\bf (a)} The restriction of the unique smooth extension $\widetilde{Y}$ of $Y$ to $\scrim$ (see Prop. 
\ref{togroup}) belongs to $\gg_\scrim$.

{\bf (b)} $\{\exp\{t\widetilde{Y}\}\}_{t\in \bR}$ is a subgroup of $SG_\scrim$.}\\

\noindent The proof of this theorem is in the Appendix.\\
As an example consider the expanding universe $M$ with cosmological horizon associated with the metric $g_{FRW}$ (\ref{cosmo0}) 
with $\kappa=1$ and $a$ as in \nref{condag}. In this case $X:= \partial_\tau$ and there is a lot of Killing vectors $Y$ of 
$(M,g_{FRW})$
satisfying $g_{FRW}(Y,X) \to 0$ approaching $\scrim$. The most trivial ones are all of the Killing vectors of the surfaces at $\tau=$constant
with respect to the induced metric. We have here a Lie algebra generated by $6$ independent Killing vectors
$Y$ associated, respectively,
space translations and space rotations. In this case $g_{FRW}(Y,X)=0$ so that the associated  Killing vectors
$\widehat{Y}\spa\rest_\scrim$ belongs to $\gg_\scrim$.  This is not the whole story in the sharp case $a(\tau)= 
\gamma/\tau$ with $\gamma <0$ which corresponds to the expanding de Sitter spacetime. 
Indeed, in this case, there is another Killing vector $B$ of $g_{FRW}$ fulfilling $g_{FRW}(B,X) \to 0$ approaching 
$\scrim$. It is $B:= \tau\partial_\tau + r \partial_r$. $B$, extended to $M \cup \scrim$, gives rise to the structure of a 
{\em bifurcate Killing horizon} \cite{KW}.\\
A last technical result, proved in the Appendix and useful in the forthcoming discussion, is 

\proposizione \label{positiveenergy} {\em Let $(M,g, \Omega, X,\gamma)$ 
be an expanding universe with cosmological horizon and $Y$  a smooth vector field of $(M,g)$
which tends to the smooth field $\widetilde{Y} \in \gg_\scrim$ pointwisely.\\
If there is an open set $A\subset \hM$ with $A \supset \scrim$ where $Y\spa\rest_{A \cap M}$ is timelike and 
future directed, then,  everywhere on $\scrim$,
\beq
\widetilde{Y}(\ell,s) = f(s) \partial_\ell\:, \quad \mbox{for some $f\in C^\infty(\bS^2)$, with $f(s)\geq 0$ on $\bS^2$.}
\eeq}
\section{Preferred states induced by the cosmological horizon.}
In this section $(M,g, \Omega, X,\gamma)$ is an expanding universe with cosmological horizon. 
Since $(M,g)$ is globally hyperbolic per definition, one can
study properties of quantum fields propagating therein, following the algebraic 
approach in the form presented in \cite{KW,Wald2}.

\ssb{QFT in the bulk} Consider real linear bosonic QFT in $(M,g)$ 
based on the symplectic space $(\cS(M),\sigma_M)$, where $\cS(M)$ is the space of real
smooth, compactly supported on Cauchy surfaces, solutions $\varphi$ of 
\beq 
P\varphi =0\:, \quad \mbox{where $P$ is the Klein-Gordon operator  $P=\Box  + \xi R  + m^2$}\:.\label{PKG}
\eeq
with $\Box = -\nabla_a \nabla^a$, $m>0$ and $\xi\in \bR$ constants.
The nondegenerate, Cauchy-surface independent, symplectic form $\sigma_M$ is:
\beq \displaystyle \sigma_M(\varphi_1,\varphi_2) := \int_S \left(\varphi_2 \nabla_N \varphi_1 - \varphi_1 
\nabla_N \varphi_2\right)\: d\mu^{(S)}_g \quad\forall\varphi_1, \varphi_2\in\sS(M)\:, \label{sigmaM}\eeq 
$S$ being any Cauchy surface of $M$ with normal unit future-directed vector $N$ and $3$-volume measure 
$d\mu^{(S)}_g$ induced by $g$.  
As is well known \cite{BR,BR2}, it is possible to associate canonically any symplectic space, for instance  
$(\sS(M),\sigma_M)$, a {\bf Weyl $C^*$-algebra}, $\cW(M)$ in this case. This is, up to (isometric) 
$*$-isomorphisms, unique and its generators $W_M(\varphi)\neq 0$, $\varphi \in \sS(M)$, satisfy {\bf Weyl 
commutation relations} (from now on we employ conventions as in \cite{Wald2})
 \beq W_M(-\varphi)= W_M(\varphi)^*\:,\quad\quad W_M(\varphi)W_M(\varphi') = 
 e^{i\sigma_M(\varphi,\varphi')/2} W(\varphi+\varphi')\:.\label{Weylc}\eeq 
$\cW(M)$  represents
the basic set of quantum observable associated with the bosonic field $\phi$ propagating in the bulk 
spacetime $(M,g)$. 

The main goal of this section is to prove that the geometric structures on $(M,g, \Omega, X,\gamma)$ pick out a very remarkable algebraic state $\omega$
on $\cW(M)$, which, among other properties turns out to be invariant under the natural action of every Killing isometry of $(M,g)$ 
which preserves $\scrim$. This happens provided a certain algebraic interplay between QFT in $M$ and QFT on $\scrim$ exists.
\ssb{Bosonic QFT on $\scrim$ and $SG_\scrim$-invariant states} Referring to $\scrim \equiv \bR \times \bS^2$, 
consider 
\beq \sS(\scri) := \left\{ \left.\psi\in C^\infty(\bR \times \bS^2)  \: \:\right|\:\: \psi\:, \partial_\ell 
\psi \in  L^2(\bR\times \bS^2, d\ell \wedge \epsilon_{\bS^2}(\theta,\phi)\right\} \:, 
\eeq
$\epsilon_{\bS^2}$ being the standard volume form of the unit $2$-sphere,
and the nondegenerate symplectic form $\sigma$ 
\beq \sigma(\psi_1,\psi_2) := \int_{\bR\times \bS^2} 
\left(\psi_2 \frac{\partial\psi_1}{\partial \ell}  - 
\psi_1 \frac{\partial\psi_2}{\partial \ell}\right) d\ell \wedge \epsilon_{\bS^2}(\theta,\phi)\quad\;\forall
\psi_1,\psi_2 \in \sS(\scri)\label{oldsigma}\:.
\eeq
As in the previous section, we associate to $(\sS(\scrim),\sigma)$ the $C^*$-algebra $\cW(\scrim)$ whose 
generators $W(\psi)\neq 0$ satisfy the Weyl commutation relations \eqref{Weylc}
\remark \label{Gscriminvariance} Exploiting the given definitions, it is straightforwardly proved that $(\sS
(\scri),\sigma)$ is {\em invariant under the pull-back action of  $SG_\scrim$}. In other words 
 (i) $\psi \circ g \in \sS(\scrim)$ if $\psi \in \sS(\scrim)$ and also (ii) $\sigma(\psi_1 \circ g,\psi_2 \circ g) = \sigma(\psi_1,\psi_2)$ for all $g\in SG_\scrim$
 and $\psi_1,\psi_2\in \sS(\scrim)$.  
As a well known consequence \cite{BR2,BGP}, $SG_\scrim$ induces a $*$-automorphism $G_{\scrim}$-representation
 $\alpha : \cW(\scrim) \to \cW(\scrim)$, uniquely individuated by linearity and continuity 
  by  the requirement
\beq 
\alpha_g(W(\psi)):= W(\psi\circ g^{-1})\:,\quad \mbox{$\psi \in \sS(\scrim)$ and $g\in G_{\scrim}$}
\label{alpha}.
\eeq

\noindent Since we are interested in  physical properties which are $SG_\scrim$-invariant, we face the issue about the existence 
of $\alpha_g$-invariant algebraic states on $\cW(\scrim)$ with $g\in SG_\scrim$.\\ 

We adopt here the definition of {\bf quasifree state} given in \cite{KW}, and also adopted in
\cite{DMP,M1,M2}. Consider the quasifree state $\lambda$ defined on $\cW(\sS(\scrim))$ 
  unambiguously defined as follows: if $\psi,\psi' \in \sS(\scrim)$, then
\beq 
\lambda(W(\psi)) = e^{-\mu(\psi,\psi)/2}\:, \quad \mu(\psi,\psi'):=  
Re \int_{\bR\times \bS^2} \sp\sp 2k\Theta(k)\overline{\widehat{\psi}(k,\theta,\phi)} \widehat{\psi'}(k,\theta,\phi) dk\wedge
 \epsilon_{\bS^2}(\theta,\phi) \:,
 \label{statescrim}\eeq
the bar denoting 
the complex conjugation, $\Theta(k):=0$ for $k< 0$ and $\Theta(k) :=1$ for $k\geq0$; here
we have used the $\ell$-Fourier-Plancherel transform $\widehat{\psi}$ of $\psi$:
\beq \widehat{\psi}(k,\theta,\phi) := \int_{\bR} \frac{e^{i k\ell}}{\sqrt{2\pi}} \psi(\ell,\theta,\phi) 
d\ell\:,
\quad (k,\theta,\phi)\in \bR \times \bS^2\:.\label{Aadded}\eeq
The constraint 
\beq
|\sigma(\psi,\psi')|^2 \leq 4\:\mu(\psi,\psi)\mu(\psi',\psi')\:, \quad\quad \mbox{for every $\psi,\psi' \in 
\sS$}\label{sm0}\:,
\eeq
which must hold for every quasifree state (see Appendix A in \cite{M1}), is  
fulfilled by the scalar product $\mu$, as the reader can verify by inspection exploiting \eqref{Aadded} and 
the definition of $\sigma$ Consider the GNS representation of $\lambda$, $(\gH, \Pi, \Upsilon)$. Since 
$\lambda$ is quasifree, $\gH$ is a bosonic Fock space $\cF_+(\cH)$ with cyclic vector $\Upsilon$ given by 
the Fock vacuum and $1$-particle Hilbert $\cH$ space obtained as the Hilbert completion of the complex space 
generated by the ``positive-frequency parts'' $\Theta\widehat{\psi}=: K_{\mu}\psi$, of every wavefunction 
$\psi\in\sS(\scrim)$, with  the scalar product $\langle \cdot, \cdot\rangle$ individuated by $\mu$, as stated 
in (ii) of Lemma A1 in the Appendix A of \cite{M1}. In our case 
\beq\label{scalar}
\langle
K_\mu\psi,K_\mu\psi^\prime\rangle=\int\limits_{\bR\times\bS^2}2k\Theta(k)\overline{\widehat{\psi}(k,
\theta,\phi)}\widehat{\psi^\prime}(k,\theta,\phi)dk\wedge\epsilon_{\bS^2}(\theta,\phi).
\eeq
\indent The map  $K_{\mu}: \sS(\scrim) \to \cH$ is $\bR$-linear and has a dense complexified range. 
  A state similar
 to $\lambda$, and denoted by the same symbol, has been defined on $\scri\simeq \bR \times \bS^2$
 in \cite{DMP,M1,M2}\footnote{In \cite{DMP,M1} a different, but unitarily-equivalent, 
Hilbert space representation was used referring to the measure $dk$ instead of $2kdk$.
Features of Fourier-Plancherel theory on $\bR \times \bS^2$ were discussed in the Appendix C of \cite{M2}.}
and, barring minor adaption, it enjoys exactly the form \eqref{statescrim}. Therefore, we can make use of 
Theorem 2.12 in \cite{DMP} we know that $\la$ {\em is pure}. Furthermore the one-particle space 
 $\cH$ of its GNS representation is isomorphic to the separable Hilbert
  space $L^2(\bR^+\times \bS^2; 2kdk \wedge \epsilon_{\bS^2})$.
%
%
%
%
%
 
The state $\lambda$ enjoys further remarkable properties in reference to the group $SG_{\scrim}$.
Particularly, since $(\gH,\Pi,\Upsilon)$ is its GNS triple, {\em $\lambda$ turns out to be invariant under the
$*$-automorphisms representation \eqref{alpha} for all $g\in SG_{\scrim}$}. In other words $\lambda(\alpha_g(A))
$ turns out to be equal to $\lambda(a)$ for all $A \in\cW(\scrim)$ and for all $g\in SG_\scrim$ as it can be 
realized out of the straightforward extension to the whole algebra of the 
the following unitary action $V$ of $SG_{\scrim}$ on the one-particle Hilbert space $\cH$:
\beq
\left(V_{(R,a,b)} {\varphi}\right)(k,s) :=  e^{a(R^{-1}(s))} e^{-ikb(R^{-1}(s))} 
{\varphi}\left( e^{a(R^{-1}(s))} k, R^{-1}(s)\right) \quad \mbox{for all $\varphi\in \cH$} \label{V}\:,
\eeq
being $g = (R,a,b) \in SG_\scrim$ and $s=(\theta,\phi)$.
Furthermore, by standard manipulation, one can realize that the unique 
unitary representation $U: SG_\scrim \ni g \mapsto U_g$ that implements $\alpha$ in $\gH$ while leaving 
$\Upsilon$ invariant, preserves $\cH$ and it is unambiguously determined by $U\spa\rest_\cH$. $U$ has the
following tensorialised form 
\beq U= I \oplus U\spa\rest_\cH \oplus (U\spa\rest_\cH \otimes \:U\spa\rest_\cH) \oplus 
(U\spa\rest_\cH \otimes \:U\spa\rest_\cH \otimes \:U\spa\rest_\cH) \oplus
\cdots\label{tensor}
\eeq
Finally the restriction of $U$ on the one-particle Hilbert space $\cH$ is an irreducible representation.

A second important result concerns the positive-energy/uniqueness properties of $\lambda$.
In Minkowski QFT positivity of energy, is a stability requirement and in general spacetimes the 
notion of energy is associated to that of a Killing time. This interpretation can be extended to this 
case too, namely to the theory on $\scrim$. The positive-energy 
requirement is fulfilled for the ``asymptotic'' notion of time associated to the limit values $\widetilde{Y}$ 
towards $\scrim$  of a timelike future-directed vector field $Y$ in $M$, when $\widetilde{Y} \in \gg_\scrim$.
Notice that $Y$ may not be a Killing vector outside $\scrim$; it is enough that $Y \to \widetilde{Y} \in \gg_
\scrim$. This includes the case $Y=X$ in particular, due to Proposition \ref{X}.\\
In the following, $\{\exp\{tZ\}\}_{t\in \bR}$ is 
the one-parameter subgroup of $G_{\scrim}$ generated by any $Z\in \gg_\scrim$  and $\{\alpha^{(Z)}_t\}_{t\in 
\bR}$ is the associated one-parameter group of $*$-automorphisms of $\cW(\scrim)$ (\ref{alpha}).

\proposizione \label{PP} {\em Consider an expanding universe with cosmological horizon $(M,g,X,\Omega,\gamma)$, 
the quasifree, pure, $SG_{\scrim}$-invariant  state $\lambda$ on $\cW(\scrim)$ defined in (\ref{statescrim}) and a timelike future-directed vector field $Y$ in $M$ such that
 $Y \to \widetilde{Y}\in \gg_\scrim$ pointwisely approaching $\scrim$ ($Y=X$ in particular, in view of Proposition \ref{X}). The following holds.\\
{\bf (a)} The unitary group $\{U^{(\widetilde{Y})}_t\}_{t\in \bR}$ which implements $\alpha^{(\widetilde{Y})}$ 
leaving fixed the cyclic GNS vector in the GNS representation of $\lambda$
is strongly continuous with nonnegative self-adjoint generator $H^{(\widetilde{Y})} = 
-i\frac{d}{dt}_sU^{(\widetilde{Y})}_t|_{t=0}$.\\
{\bf (b)} The restriction of $H^{(\widetilde{Y})}$ to the one-particle space has no zero 
modes if and only if $\widetilde{Y}$ vanishes on 
a zero-measure subset of $\scrim$ }.\\

\noindent{\em Proof}. From Proposition \ref{positiveenergy} one has that $\widetilde{Y}(\ell,s)=
f(s)\partial_\ell$ for some non negative smooth function $f:\bS^2\to \bR$. Therefore $\exp\{t\widehat{Y}\}$
amounts to the  displacement $(\ell,s) \to (\ell+f(s)t,s)$. As a consequence of the previous discussion, the one parameter 
group $\alpha^{(\widetilde{Y})}$ is unitarily represented by $\{U^{(\widetilde{Y})}_t\}_{t\in \bR}$. 
$U^{(\widetilde{Y})}_t$ is the tensorialisation (as in (\ref{tensor})) of the (representation of the) unitary 
group in the one-particle space $V_t: \cH \to \cH$, with
$$(V_t \phi)(k,s) = e^{itk f(s)}\psi(k,s) = \left(e^{ith^{(\widetilde{Y})}}\psi\right)(k,s)\:,\quad \mbox{for all $\phi \in \cH$.}$$
  From standard theorems of operator theory one obtains that $\bR \ni t \mapsto V_t$
is strongly continuous with self-adjoint generator $h^{(\widetilde{Y})}$, in the one-particle space $\cH=L^2(
\bR^+\times \bS^2; 2kdk\wedge \epsilon_{\bS^2})$, given by $(h^{(\widetilde{Y})} {\phi})(k,s) = k f(s){\phi}(
k,s)$, defined in the dense domains $\cD(h^{(\widetilde{Y})})$ made of  the elements of the
Hilbert space $L^2(\bR^+\times \bS^2; 2kdk\wedge \epsilon_{\bS^2})$
such that the right-hand side belongs to $L^2(\bR^+\times \bS^2; 2kdk\wedge \epsilon_{\bS^2})$. It is so 
evident that, since $f\geq0$, for every $\psi \in \cD(H)$
\beq  \langle \phi,h^{(\widetilde{Y})} \phi \rangle = \int_0^{+\infty} 2kdk \int_{\bS^2} \epsilon_{\bS^2}(s) |\phi(k,s)|^2 k f(s) \geq 0\:,\label{zero}\eeq
and thus $\sigma(h^{(\widetilde{Y})})\subset [0,+\infty)$. Passing to the whole Fock space by (\ref{tensor}) 
the result remains unchanged for the whole generator $H^{(\widetilde{Y})}= 0+ h^{(\widetilde{Y})} \oplus I
\otimes h^{(\widetilde{Y})} \oplus h^{(\widetilde{Y})}\otimes  I\oplus \cdots$ using standard properties of 
generators. The last statement is a trivial consequence of (\ref{zero}) using $\widetilde{Y}= f \partial_
\ell$. $\Box$\\

\noindent  The result applies in particular for $\widetilde{Y}= \partial_\ell$, since
 it is always possible to view $\partial_\ell$ as the limit value of some timelike vector field of $M$.
For expanding universes with cosmological horizon as described in section \ref{FRWsection},
if  $X := -\gamma
\partial_\tau$, then $X \to \partial_\ell$ while approaching $\scrim$. In this above case the energy-positivity 
property applies for $X$ and there are no zero modes.\\
This is not the whole story, since the positive-energy property for $\partial_\ell$, determines completely 
$\lambda$.  

\teorema  \label{teoremauniqueness}  {\em Consider the state $\lambda$ defined in (\ref{statescrim}) and its GNS representation.  The following holds.\\
{\bf (a)}  The state $\lambda$  is the unique pure quasifree state
 on $\cW(\scrim)$ satisfying both:

(i) it is invariant 
under $\alpha^{(\partial_\ell)}$,

(ii) the unitary group which implements $\alpha^{(\partial_\ell)}$ leaving fixed the cyclic GNS vector 
is strongly continuous with nonnegative self-adjoint generator (energy positivity condition).\\
 {\bf (b)}  Each folium of states on $\cW(\scrim)$ contains at most one pure $\alpha^{(\partial_\ell)}$-invariant state.}\\

\noindent{\em Proof}. 
The proofs of (a) and (b), though rather technical, are identical to those of the corresponding
statements in Theorem 3.1 of \cite{M1}, where, in the cited proof, $\cF$ refers to a Bondi frame. 
This holds since the self-adjoint generator of the unitary group 
$t \mapsto U_t$, implementing $\{\alpha^{(\partial_\ell)}_t\}_{t\in \bR}$ and leaving $\Upsilon$ invariant,
is the tensorialisation of the positive self-adjoint generator $H$ acting in the one-particle space 
$L^2(\bR^+\times \bS^2; 2kdk\wedge \epsilon_{\bS^2})$ as $(H \widehat{\psi})(k,\theta,\phi) = k 
\widehat{\psi}(k,\theta,\phi)$. Note that $H$ is defined in 
the dense domains of  the elements of the Hilbert space $L^2(\bR^+\times \bS^2; 2kdk\wedge \epsilon_{\bS^2})$
such that the right-hand side is still in $L^2(\bR^+\times \bS^2; 2kdk\wedge \epsilon_{\bS^2})$. Hence 
$\sigma(H) = \sigma_c(H) =[0,+\infty)$.\\
The action of the one-parameter subgroup $\bR \ni t \mapsto g^{(\partial_\ell)}(t)$ of $G_{\scrim}$ on fields
defined on $\scrim$ coincides exactly with the one-parameter subgroup of the $BMS$ group on fields defined on
$\scrim$. Furthermore also the unitary representations 
of $SG_{\scrim}$ and of the BMS group are identical when restricted to those subgroups. $\Box$

\ssb{Interplay of QFT in $M$ and QFT on $\scrim$}
While in the previous section we have shown that it exists a preferred quasifree pure state $\lambda$ invariant under the 
action of $SG_\scrim$ and enjoying some uniqueness properties, we wonder now if it is possible to induce  
a state $\lambda_M$ on the algebra of field observables in the bulk starting from $\lambda$. If this is the case, we would expect $\lambda_M$ to
fulfil some invariance properties with respect to the possible isometries individuated by Killing vectors 
which preserve $\scrim$. 
To this avail, we concentrate beforehand on algebraic properties, establishing the existence of a nice 
interplay between $\cW(\scrim)$ and $\cW(M)$ under suitable hypotheses on the considered symplectic forms. 
That interplay will be used to define $\lambda_M$ in the next subsection.\\
The symplectic form $\sigma_M$ on $\cS(M)$ defined in (\ref{sigmaM}) can be equivalently rewritten as the 
integral of a $3$-form,
\beq \displaystyle \sigma_M(\varphi_1,\varphi_2) :=\int_S  \chi(\varphi_1,\varphi_2) \label{sigmaM2} 
=\int_S\frac{1}{6}\left(\varphi_1 \nabla^\mu \varphi_2 - \varphi_2 \nabla^\mu \varphi_1\right)
\:\sqrt{-\widehat{g}}\: \epsilon_{\mu\alpha\beta\gamma}\: dx^\alpha\wedge dx^\beta \wedge dx^\gamma  \:,
\eeq
where $\epsilon_{\mu\alpha\beta\gamma}$ is the totally antisymmetric Levi Civita symbol, $S$ is a future
oriented Cauchy surface and the second equality holds in any  local coordinate patch.
  
Notice that, even though $S$ is moved back in the past and it seems to tend to coincide with $\scrim$, this 
is not necessarily the case, since $\scrim$ and Cauchy surfaces in $M$
may have different topologies. In particular, information could get lost through the time-like past infinity 
$i^-$, the tip of the cone representing $\scrim$. That point does not belong to $\hM$ in our hypotheses.
However  one may expect that, in certain cases at least, assuming that each $\varphi_i$ extends to $\Gamma 
\varphi_i \in \sS(\scrim)$ smoothly, it holds
\beq
\sigma_M(\varphi_1,\varphi_2) = \int_\scrim  \chi(\Gamma\varphi_1,\Gamma\varphi_2)\:. 
\eeq
Now, by direct inspection one verifies that, for $\psi_1,\psi_2 \in \sS(\scrim)$,
\beq \int_\scrim  \chi(\psi_1,\psi_2) = \gamma^2\int_{\bR\times \bS^2} 
\left(\psi_2 \frac{\partial\psi_1}{\partial \ell}  - 
\psi_1 \frac{\partial\psi_2}{\partial \ell}\right) d\ell \wedge \epsilon_{\bS^2}(\theta,\phi)\:,\eeq
where $\gamma$ is the last constant in $(M,g,\Omega,X,\gamma)$.
Following this way one is led to expect that
\beq 
\sigma_M(\varphi_1,\varphi_2) =  \sigma(\gamma\Gamma\varphi_1,\gamma\Gamma\varphi_2) \:.\label{idsym} \eeq
Notice that this result is by no means trivial and it might not hold, since it strictly depends on the 
behaviour of the solutions of Klein-Gordon equations across $\scrim$. 

Here we investigate the consequences of (\ref{idsym}) under the hypothesis that such an identity holds true.
The existence  of $\Gamma: \cS(M)\to \sS(\scrim)$ fulfilling (\ref{idsym}) implies the 
existence of a isometric $*$-homomorphism $\imath : \cW(M)\to \cW(\scrim)$. In this way the field 
observables of the bulk are mapped into observables of the theory on $\scrim$. Moreover, the state $\lambda$ 
on $\scrim$ induces a preferred state $\lambda_M$ on $\cW(M)$ via pull-back. This state enjoys interesting 
invariance properties with respect to the symmetries of $(M,g)$ which preserve $\scrim$, as well as a 
positivity property with respect to timelike Killing vectors of $M$ which preserve $\scrim$.

\teorema \label{theoremimath}{Consider an expanding universe with cosmological horizon $(M,g,X,\Omega,\gamma)$ 
and suppose that every $\varphi \in \cS(M)$
extends smoothly to some $\Gamma \phi \in \sS(\scrim)$ in order that (\ref{idsym}) holds true:
\beq
\sigma_M(\varphi_1,\varphi_2) = \sigma(\gamma\Gamma\varphi_1,\gamma\Gamma\varphi_2)\:, 
\quad \mbox{for every $\varphi_1,\varphi_2 \in \cS(M)$.}\nonumber
\eeq
In these hypotheses, there is an (isometric) $*$-homomorphism $\imath : \cW(M) \to \cW (\scrim)$ that identifies the Weyl 
$C^*$-algebra of the bulk $M$ with a sub $C^*$-algebra of the boundary $\scrim$; it is completely 
determined by the requirement:
\beq
\imath\left( W_M(\varphi)\right):= W(\gamma\Gamma \varphi)\:, \quad \mbox{for all $\varphi \in \cW(M)$.} \label{imath} 
\eeq}

\noindent {\em Proof}. 
Notice that the linear map $\gamma\Gamma : \cS(M)\to \sS(\scrim)$
has to  be injective due to  nondegenerateness of $\sigma$ and (\ref{idsym}).
Consider the sub Weyl-$C^*$-algebra $A_M$ of $\cW(\scrim)$ generated by the elements   
$W(\gamma\Gamma \varphi)$ with $\varphi \in \cS(M)$. Since Weyl $C^*$-algebras are determined up to 
(isometric) $*$-algebra isomorphisms, $A_M$  is nothing but the Weyl $C^*$-algebra associated with the symplectic space 
$(\gamma\Gamma(\cS(M)), \sigma)$ and the map $\gamma\Gamma : \cS(M) \to \Gamma(\cS(M))$ is an isomorphism of symplectic spaces.
Under these hypotheses \cite{BR2}, there is a unique (isometric)  $*$-isomorphism $\imath : \cW(M) \to A_M\subset 
 \cW (\scrim)$ 
completely individuated by (\ref{imath}). $\Box$\\

\ssb{The preferred invariant state $\lambda_M$} We proceed to show that, in the hypotheses of Theorem 
\ref{theoremimath},
a preferred state $\lambda_M$ on $\cW(M)$ is induced by $\lambda$. That state enjoys very remarkable physical properties.\\
 From now on, if $Y$ is a complete Killing vector  of $(M,g)$, the associated one-parameter group of $g$-isometries,
$\{\exp\{tY\}\}_{t\in \bR}$, preserves under pull-back action $\sigma_M$. Hence \cite{BR2,BGP} there is a 
unique isometric $*$-isomorphism $\beta^{(Y)}_t: \cW(M) \to \cW(M)$ induced by 
$$\beta^{(Y)}_t(W_M(\varphi)) := W_M(\varphi \circ \exp\{-tY\})\:, \quad\mbox{for every 
$\varphi\in \cS(M)$.}$$ In the following we shall call $\beta^{(Y)}:= \{\beta^{(Y)}_t\}_{t\in
\bR}$ the {\bf natural $*$-isomorphism action of 
$\{\exp\{tY\}\}_{t\in \bR}$ on $\cW(M)$}. Similarly, every 
$Z \in \gg_\scrim$ has a {\bf natural action $\alpha^{(Z)}$ on $\cW(\scrim)$} in terms of isometric $*$-isomorphism, 
obtained by requiring,
$$\alpha^{(Z)}_t(W(\psi)) := W(\psi \circ \exp\{-tZ\})\:, \quad\mbox{for every 
$\psi\in \sS(\scrim)$,}$$
since the pull-back action of $\{\exp\{tZ\}\}_{t\in \bR}$, generated by $Z$ on fields of $\sS(\scrim)$ 
preserves $\sigma$.\\
To stress a further important point, let us consider an expanding universe with cosmological horizon 
$(M,g,X,\Omega,\gamma)$ and let us suppose that every $\varphi \in \cS(M)$
extends smoothly to some $\Gamma \varphi \in \sS(\scrim)$ in order that (\ref{idsym}) holds true.
In this case there is a uniquely defined smooth function $\widehat{\varphi}$ defined on $M\cup \scrim$, that reduces to 
$\varphi$ in $M$ and to 
$\Gamma \varphi$ on $\scrim$. If $Y$ is a complete Killing vector of $(M,g)$ preserving $\scrim$, 
the one parameter group generated by its unique extension $\widehat{Y}$ to $M\cup \scrim$ (Proposition \ref{togroup} and Theorem  \ref{theorem2}) 
acts on $\widehat{\varphi}$ globally. 
Taking the relevant restrictions of scalar fields and Killing vector fields we obtain:
\beq
(\Gamma\varphi)\circ\exp\{t\widetilde{Y}\} = 
\Gamma\left(\varphi \circ \exp\{t{Y}\}\right)\:,
\eeq
where, as usual, $\widetilde{Y} := \widehat{Y}\spa\rest_\scrim$.
As a straightforward consequence it holds
\beq
\imath\left(\beta^{(Y)}_t (a)\right) =\alpha^{(\widetilde{Y})}_t (\imath(a))\:, 
\quad \mbox{for all $a\in \cW(M)$ and $t\in \bR$}\label{central}\:.
\eeq

\teorema{Consider an expanding universe with cosmological horizon $(M,g,X,\Omega,\gamma)$ fulfilling the 
 hypotheses of Theorem \ref{theoremimath}. Let $\lambda_M : \cW(M) \to \bC$ be the state
induced by $\lambda$ defined in (\ref{statescrim}) through the isometric $*$-homomorphism $\imath$ 
(\ref{imath}):
\beq
\lambda_M(a) := \lambda(\imath(a))\:, \quad \mbox{for all $a\in \cW(M)$.} \label{lambdaM}
\eeq
$\lambda_M$ enjoys the following properties:\\
{\bf (a)} Whenever $(M,g)$ admits some complete Killing vector field $Y$ preserving $\scrim$, then letting 
$\beta^{(Y)}$ be the natural action on $\cW(M)$, $\lambda_M$ is invariant under $\beta^{(Y)}$ and the unitary 
one-parameter group 
$\{U^{(Y)}_t\}_{t\in \bR}$, which implements $\beta^{(Y)}$ in the GNS representation of $\lambda_M$ leaving fixed the cyclic vector, 
is strongly continuous.\\
{\bf (b)} If $Y$ above is everywhere timelike and future-directed in $M$, then
(i) the one-parameter group $\{U^{(Y)}_t\}_{t\in \bR}$ has 
positive self-adjoint generator, 
(ii) that generator has no zero-modes in the one-particle subspace, if $\widetilde{Y}=0$ on a zero-measure 
subset of $\scrim$.\label{theoremlambdaM}}

\remark  As noticed before Proposition \ref{PP}, positivity of energy is a stability requirement. The 
statement (b) of the theorem assures that, in the presence of a timelike Killing vector out of which defining
the notion of energy, if it preserves $\scrim$, the condition of energy positivity holds true. If such a
timelike Killing vector is absent, then Proposition \ref{PP} assures nonetheless the validity of a 
positivity-energy  condition, particularly with respect to the conformal Killing vector $X$. 

\ssb{Testing the construction for the de Sitter case and for other FRW metrics} We proceed to show that 
the hypotheses of Theorem \ref{theoremimath} are valid when $(M,g,X,\Omega,\gamma)$ is in the class of the 
FRW metrics considered in section \ref{FRWsection}, so that the preferred state $\lambda_M$ exists for those 
spacetimes. That class includes the expanding region of de Sitter spacetime (see \cite{Bros, Bros2} for a
related analysis in the framework of Wightman's axioms). We shall verify, in this last 
case, that the preferred state $\lambda_M$ is nothing but the well-known {\em de Sitter Euclidean vacuum} or 
{\em Bunch-Davies state}, $\omega_E$ \cite{SS, BD, Allen}. 
Let us start with de Sitter scenario. The expanding de Sitter region is 
\beq\label{dSexp}
M \simeq (-\infty,0) \times \bR^3\:,  \quad g=a^2(\tau)\left[-d\tau\otimes d\tau + dr\otimes dr+r^2d
\bS^2(\theta,\varphi)\right], 
\eeq
where $\tau \in (-\infty,0)$ and where $r,\theta,\phi$ are standard spherical coordinates on 
$\bR^3$, whereas $a(\tau) = \gamma/\tau$ for some constant  $\gamma<0$, so that  and  $R=12/\gamma^2$. 
A class of, generally complex, solutions $\Phi_{\bk}$, $\bk\in \bR^3$ of \eqref{PKG} is\footnote{The form of
the modes as presented in \cite{BD,BiD} is different both since in \cite{SS, BD} the contracting region of de 
Sitter spacetime was considered and due to the absence of the overall exponential $\exp -i\pi \nu/2$, which 
would affect the final results and the normalisation (\ref{norm}) for $\nu$ imaginary, but not the final form
of the two-point function.}
\beq
\Phi_\bk(\tau,\bx):= \frac{e^{i\bk\cdot \bx}}{(2\pi)^{3/2}} \frac{\chi_\bk(\tau)\label{ph}}{a(\tau)}\:,
\label{Phik}
\eeq
where, according to \cite{SS}, it holds
\beq
\chi_\bk(\tau) := \frac{1}{2} \sqrt{-\pi\tau}\: e^{i\pi \nu/2} \overline{H^{(2)}_\nu(-k\tau)}\:,
 \quad \mbox{where}\quad  \nu := \sqrt{\frac{9}{4}-12(m^2 R^{-1} + \xi)}\:\label{chi}
\eeq
being $k:= |\bk|$ and $H^{(2)}_\nu$ is the second-type Hankel function. The sign in front of the square root in the definition of 
$\nu$ (which may be imaginary) does not affect the right-hand side of (\ref{chi})  and it could be fixed 
arbitrarily (either for $\nu$ real or imaginary). With these choices one finds the time-independent 
normalisation 
 \beq
\frac{d \overline{\chi_\bk(\tau)}}{d\tau}  \chi_\bk(\tau)- \overline{\chi_\bk(\tau)}
\frac{d \chi_\bk(\tau)}{d\tau}= i \:,\quad \mbox{for all $\tau\in (-\infty,0)$.} \label{norm}
\eeq

\noindent Let us now show how $\omega_E$ is defined.
 To this end, take any $\ph \in \cS(M)$ and a Cauchy surface $\Sigma_{\tau}$ in $(M,
 g)$ at fixed $\tau$. Define
 \beq
\widetilde{\ph}(\bk) := -i \int_{\bR^3} \left[ \frac{\partial \overline{\Phi_\bk(\tau,\bx)}}{\partial\tau}\ph(\tau,\bx)  - 
 \overline{\Phi_\bk(\tau,\bx)} \frac{\partial \ph(\tau, \bx)}{\partial\tau} \right] a(\tau)^2 d\bx, \:. \label{FF}
\eeq
where, per direct inspection, the right-hand side of (\ref{FF}) does not depend on the choice of $\tau$.
 
Furthermore, $H_\nu^{(2)}(z)$ decays as $z^{-1/2}$ as $|z|\to \infty$, $\widetilde{\ph}\in C^{\infty}(\bR^3
\setminus\{0\})$ and it vanishes for $|\bk|\to \infty$ faster than every power $|\bk|^{-n}$, $n\in\bN$. From 
the known behaviour of the functions $H^{(2)}_\nu(z)$ in a neighbourhood of $z=0$ \cite{Grad}, one sees both 
that the leading divergence as $\bk\to 0$ due to the functions $\chi_\bk$ is of order $|\bk|^{-|Re \nu|}$ and
that $|\widetilde{\ph}|^2$, as well as $|\widetilde{\ph}|$, is integrable with respect to $d\bk$ whenever 
$|Re \nu|<3/2$ or, equivalently, $m^2 + \xi R > 0$.
Once one constructs $\widetilde{\ph}$ out of (\ref{FF}), then $\ph$ is 
\beq
\ph(\tau,\bx) = \int_{\bR^3} \left[ \Phi_{\bk}(\tau,\bx) \widetilde{\ph}(\bk)
+  \overline{\Phi_{\bk}(\tau,\bx)} \overline{\widetilde{\ph}(\bk)}
\right] \: d\bk\:. \label{FFinv}
\eeq 
This holds out of (\ref{Phik}), (\ref{norm}),  (\ref{FF}), and of the properties of Fourier transform for 
functions in $C^\infty_0(\bR^3)$.\\ 
Since when $m^2 + \xi R > 0$ and $\ph \in \cS(M)$, $\widetilde{\ph}\in L^2(\bR^3; d\bk)\cap 
L^1(\bR^3; d\bk)$ then
\beq
  -2 Im \left\{ \int_{\bR^3} \overline{\widetilde{\ph}_1}(\bk) \widetilde{\ph}_2  (\bk) d\bk\right\}
  = \int_{\bR^3} (\ph_2 \partial_\tau \ph_1 -  \ph_1 \partial_\tau \ph_2 ) \: a^2(\tau) d\bx =: 
  \sigma_M(\ph_1,\ph_2)\quad\forall\ph_1,\ph_2 \in \cS(M) \:. \label{ImsigmaM}
\eeq
The (restriction to $M$ of the) {\bf Euclidean vacuum} in de Sitter space is nothing but the quasifree state $\omega_E$
on $\cW(M)$ completely identified by
   \beq
  \omega_E(W_M(\ph)) = e^{-\frac{1}{2}  \int_{\bR^3}  \overline{\widetilde{\ph}(\bk)}\widetilde{\ph}(\bk) \: d\bk }\:,
   \quad \mbox{for every $\ph\in \cS(M)$.}   \label{OmegaE}
   \eeq
   Notice that the constraint (\ref{sm0}) is automatically fulfilled in view of (\ref{ImsigmaM}).
  
\remark 
The maximally extended de Sitter spacetime can be realized by glueing together
 two isometric spacetimes -- one expanding and the other contracting, when moving towards the future -- on 
 the common cosmological horizon. The obtained spacetime is maximally symmetric and admits $SO(1,5)$ as
  group of isometries. The state $\omega_E$  extends to a globally defined state on the whole de 
Sitter spacetime \cite{Allen} and such a state is $O(1,5)$-invariant, hence it is invariant also 
under symmetries which do not preserve the horizon. 

\teorema\label{Nicola} {\em Consider the expanding universe $(M,g,X,\Omega,\gamma)$ given by \eqref{dSexp}
with $a(\tau) = \gamma/\tau$. 
Consider a quantum scalar Klein-Gordon field propagating in $(M,g)$ with  $m^2 + \xi R > 0$. Then,\\
{\bf (a)} If $m^2 + \xi R >\frac{5}{48}R$ (see also Remark \ref{lastremark}), every $\varphi \in \cS(M)$ 
extends smoothly to some $\Gamma \phi \in \sS(\scrim)$, (\ref{idsym}) holds true and \\ 
{\bf (b)} $\lambda_M$ on $\cS(M)$ coincides with the restriction to $M$ of $\omega_E$.}\\
The proof will be given in the appendix.

\remark \label{lastremark}  The requirement $m^2 + \xi R >\frac{5}{48}R$, {\it i.e.}  $|Re \nu|<1$ 
is used to assure that $\Gamma \ph\in\sS(\scrim)$ if $\ph \in \cS(M)$. Actually the requirement can be 
dropped preserving only $m^2 + \xi R >0$ if we change definition (\ref{sigmaM}) of $\sS(\scrim)$, namely
$$\sS(\scrim):= \left\{ \psi \in C^\infty(\bR \times \bS^2)\:\:\left|\:\: \int_{\bR\times \bS^2}
|\widehat{\psi}(k,\theta,\phi)|^2 |k| \: dk \wedge \epsilon_{\bS^2}(\theta,\phi) < +\infty\right. \right\}$$
where $\widehat{\psi}$ indicates the Fourier-Plancherel transform of the {\em Schwartz distribution} $\psi$
(as discussed in the Appendix C of \cite{M2}). Then the symplectic form on $\scrim$ could be defined Fourier 
transforming along the $\bR$-direction \eqref{oldsigma}. In this way, the identity (\ref{oldsigma}) would 
hold true in a weaker limit sense, employing a suitable regularisation of $\psi_1$ and or $\psi_2$ by means 
of sequences of smooth compactly supported functions. Then the construction of $\lambda$ on $\cW(\scrim)$ and 
of its GNS triple as well as the uniqueness/positive energy theorems would closely resemble to our previous
analysis. 
 
\noindent To conclude we have the last promised theorem proved in the Appendix: The hypotheses of Theorem 
\ref{theoremimath} are 
fulfilled, and thus $\lambda_M$ is defined, for FRW metrics as described in section \ref{FRWsection} with 
$a(\tau)$ as in \nref{condag}, provided the mass 
$m$ of the Klein-Gordon field and/or the constant $\xi$ are large enough.

\teorema\label{Nicola2} {\em Consider a quantum scalar Klein-Gordon field $\ph$, satisfying \eqref{PKG}
and propagating in an expanding universe $(M,g,X,\Omega,\gamma)$. Consider $a(\tau)$ as in 
\nref{condag} and with $\ddot{a}(\tau) = 2{\gamma}/{\tau^3} + O({1}/{\tau^4})\:$ in such a way that $R=12/
\gamma^2 + O(1/\tau)$, then, if 
\beq\nonumber
M \simeq (-\infty,0) \times \bR^3\:,  \quad g=a^2(\tau)\left[-d\tau\otimes d\tau + dr\otimes dr+r^2d
\bS^2(\theta,\varphi)\right],\nonumber 
\eeq
$\tau \in (-\infty,0)$ and $r,\theta,\phi$ standard spherical coordinates on $\bR^3$, $X= \partial_\tau$ and 
$\Omega=  a(\tau) = \gamma/\tau + O(1/\tau^2)$ as $\tau \to -\infty$ for some constant $\gamma<0$.), 
whenever $m^2 \gamma^2 + 12 \xi>2$, every $\varphi \in \cS(M)$ extends smoothly to some $\Gamma\phi\in\sS(
\scrim)$ and (\ref{idsym}) holds true.}

\remark
{\bf (1)} Theorem \ref{Nicola2} is also valid relaxing the hypothesis to the case $\xi=1/6$ and $m=0$. In
this case the proof is similar to 
 that of the case studied in 
\cite{DMP,M1}.\\
{\bf (2)} The validity of Hadamard property for the states $\lambda_M$ will be investigated in a forthcoming 
paper. However, a first scrutiny shows that it does hold for the states $\lambda_M$ considered in Theorem 
\ref{Nicola2} provided the two-point function of such a state is a distribution of $\cD'(M\times M)$. The 
proof is similar to the one in \cite{M2}. The distributional requirement is fulfilled if 
the functions $\Gamma \ph$, $\ph \in \cS(M)$, satisfy a suitable decay property as $\ell \to -\infty$.

\section{Conclusions and open issues.}
In this manuscript, we were able to prove that, imposing some suitable constraints on the expansion factor 
$a(t)$, the FRW background 
can be 
extended to a larger spacetime which encompasses the cosmological horizon.
%
Such structure is later generalised in definition 3.1 where we introduce a novel notion of an expanding 
universe $(M,g)$ with geodesically complete cosmological past horizon $\scrim$. It is worth to stress that, 
in the set of backgrounds we are taking into account, besides the conformal factor $\Omega$, a relevant role
is played by a future oriented timelike vector $X$ which is a conformal Killing vector for the metric
$g$. As a byproduct of these geometric properties, we were able to construct explicitly the structure of the 
subgroup $SG_\scrim$ of the isometry group of $\scrim$, {\it i.e.}, the iterated semidirect product 
$SO(3)\ltimes\left(C^\infty(\bS^2)\ltimes C^\infty(\bS^2)\right)$. Such a result suggests us that one could 
hope to readapt in this framework some of the properties of a scalar quantum field theory as discussed in \cite{DMP, M1, M2}.

In fact, using only the universal structure of $\scrim$,
we was able to select, for the theory on the horizon, a preferred state $\lambda$ which is quasi-free and pure.
 $\lambda$ is the unique state which, besides the previous properties, is also invariant under 
the action of the horizon symmetry group; actually, uniqueness for pure quasifree states on $\cW(\scrim)$ 
holds with the only hypotheses of invariance with respect to the one-parameter group generated by 
$\partial_\ell$ and a more general uniqueness property is valid as discussed in Theorem 
\ref{teoremauniqueness}. Moreover, for any future oriented timelike vector field $Y$ in the bulk such that 
it projects on the horizon to $\widetilde Y$, {\it i. e.} a generator of the Lie algebra of $SG_\scrim$, then 
the unitary group of operators implementing the action of $\widetilde Y$  on the GNS representation of 
$\lambda$ is strongly continuous with a non negative self-adjoint generator. Finally the one-particle space 
in the GNS representation of the state $\lambda$ turns out to be an irreducible representation of the group 
of horizon symmetries $SG_\scrim$.

In section 4,  we considered a generic massive scalar Klein-Gordon equation with an arbitrary coupling to 
curvature. Under the assumption that each solution of such an equation for compactly supported initial data 
projects on the horizon to a rapidly decreasing smooth function - say $\psi$ - and that such a projection 
preserves a suitable symplectic form, then we were able to draw some interesting conclusions. As a first step
 the projection map between classical fields extends also at a level of Weyl 
algebras, namely we can embed the bulk Weyl $C^*$-algebra as a $C^*$-subalgebra of the horizon counterpart.
Furthermore such an embedding between Weyl algebras can be exploited in order to pull-back $\lambda$ to a
bulk state $\lambda_M$ which is still quasi-free and invariant under the action of any bulk isometry which 
preserves the cosmological horizon. Furthermore, whenever the Killing vector is 
everywhere future oriented and timelike, than the one-parameter group of unitary operators implementing such 
an action is positive with self-adjoint generator. 

As previously mentioned these results hold true under certain hypotheses which we tested in section 4.6
where we studied the behaviour of solutions for the Klein-Gordon equation of motion with an arbitrary coupling
to curvature both in the de-Sitter and in the FRW background.  Our analysis shows -- see theorem 4.6 -- that the 
hypotheses made at the beginning of section
4, hold true at least whenever certain conditions between the relevant parameters in the
equation of motion are satisfied. In the deSitter case $\lambda_M$ coincides with the well-known Euclidean Bunch-Davies vacuum.

On the overall we feel safe to claim that the analysis we performed proves that the investigation of a 
quantum field theory in a suitable cosmological background by means of an horizon counterpart is a viable
option. Hence, as a future perspective, one would hope as a first step 
to extend the domain of applicability of theorem 4.6, and later to further discuss the properties for the
bulk state. In particular our long-term aim is to prove both that $\lambda_M$ is pure and that it
is Hadamard so that it can be used in renormalisation procedures, especially for the stress energy tensor 
\cite{Wald2,Mstress,HWstress}. Furthermore we should also 
investigate possible relations with the adiabatic states often exploited in the study of field theories on 
FRW backgrounds \cite{Junker, Luders, Olbermann, Parker}. 
Concerning the validity of Hadamard property, it holds true for  $\lambda_M$ when $M$ is deSitter spacetime 
since in this case $\lambda_M$ is the Euclidean vacuum. However, a first scrutiny shows that it does hold for 
all the states $\lambda_M$ considered in Theorem \ref{Nicola2} provided the two-point function of such a 
state is a distribution of $\cD'(M\times M)$. The proof is almost the same as that preformed in \cite{M2}.


At last but not at least, it would be interesting to extend our results to interacting fields. 
 From a physical perspective this would be the most appealing scenario since, as mentioned in the 
introduction, nowadays cosmological models are often based upon a single scalar field whose dynamic is 
governed by a non trivial potential. It could also be worth to investigate possible applications of our
results to the description of dark matter. Being weakly interacting, it is feasible to model it, at least in
a first approximation, as a free quantum scalar field on a curved background. Although here we do not address
all the above mentioned topics, we believe that this 
manuscript could be a nice first step towards this direction and we hope to discuss many if not all these 
mentioned points in a forthcoming manuscript. 

\section*{Acknowledgements.} The work of C.D. is supported by the von Humboldt Foundation and that of N.P. 
has been supported by the German DFG Research Program SFB 676. We would like to thank K. Fredenhagen and R. Brunetti for 
useful discussions.

\appendix

\section{Proof of some technical results.} 

\noindent {\bf Proof of Proposition \ref{X}}. (a) If there were a smooth extension of $X$ to $\overline{M}$ 
it would be unique by continuity, moreover, by continuity again, it would define a Killing vector for $\hg$ 
when restricting to the surface $\scrim$, because the right-hand side of (\ref{confscri}) vanishes there.
We, in fact, will prove the existence of a smooth extension to the whole $\hM$. 
Coordinates $(\ell,\Omega, \theta, \phi)$ are defined in a neighbourhood 
$U \subset \hM$ of $\scrim = \partial M$. Using the whole class of smooth curves $\gamma : t \to 
(\ell_0, t ,\theta_0, \phi_0)$ where $(\ell_0, \theta_0, \phi_0) \in \bR \times \bS^2$
are fixed arbitrarily, 
and the transport equations \cite{Geroch2,Hall}
\begin{eqnarray} \dot{\gamma}^a \widehat\nabla_a \widehat{X}_b =
\dot{\gamma}^a\left( \widehat{F}_{ab} + \frac{1}{2} \hg_{ab} \widehat{\varphi} \right), & &
 \dot{\gamma}^a \widehat\nabla_a \widehat{\varphi} = \dot{\gamma}^a \widehat{K}_a 
 \nonumber \\
\dot{\gamma}^a \widehat\nabla_a \widehat{F}_{bc} = \dot{\gamma}^a \left(\widehat{R}_{bcad}
\widehat{X}^d
 + \widehat{K}_{[b}\:\hg_{c]a}\right)
 , & &
 \dot{\gamma}^a \widehat\nabla_a \widehat{K}_b = \dot{\gamma}^a \left( \widehat{X}^d
 \widehat{\nabla}_d \widehat{L}_{ab} + \widehat{\varphi} \widehat{L}_{ab} +2 \widehat{R}_{d(a} \:\widehat{F}\
 _{b)}\:^{d}\right)
\label{OPEC}
\end{eqnarray}
(where $\widehat{L}_{ab}:= \widehat{R}_{ab} -\frac{1}{6} \hg_{ab}\widehat{R}$)
 we can ``transport'' $X$, $F_{ab}=
\widehat{\nabla}_a X_b - \widehat{\nabla}_bX_a$, $\varphi \hg := \frac{1}{2} \cL_X(\hg)$, and 
$K_a := \widehat\nabla_a \varphi$ beyond $\scrim$ in $U$.
The transported fields $\widehat{X}$, $\widehat{F}$, $\widehat{\varphi}$, and $\widehat{K}_a$ are nothing but
the solutions of the first order differential equations (\ref{OPEC}), with initial conditions given by the 
known fields ${X}$, ${F}$, $\varphi$, $K$ evaluated on a fixed smooth surface $\Omega=\Omega(\ell,\theta,\phi
)$ completely included in $M\cap U$. In $M$, $\widehat{X}$ coincides with $X$ itself (and $\widehat{F}$
coincides with $F$ itself and so on),  since every conformal Killing vector field fulfils transport
equations (\ref{OPEC}) \cite{Geroch2,Hall} and uniqueness theorem holds for solutions of ordinary 
differential equations. Outside $M$ one gets a smooth field $\widehat{X}$ anyway, due to the jointly
dependence of solution of differential equations from the initial data (assigned on a smooth surface as well
). Obviously the constructed field $\widehat{X}$ does not need to fulfil conformal Killing equations outside
$\overline{M}$.  In this way we have  constructed a smooth extension $\widehat{X}$ of $X$ on the open set  
$M \cup U$ inclosing $\scrim$, the further extension to $\hM$ is now trivial, using standard smoothing 
technology. By continuity, $\cL_{\widehat{X}} = \Omega^{-1} X(\Omega) \hg$ must hold on $\scrim$. This means
that the right-hand side smoothly extends there (to zero by hypotheses). In particular,  since  $\Omega=0$ on
$\scrim$, $\widehat{X}(\Omega)=0$ on $\scrim$. That is $\langle \widehat{X}\spa\rest_\scrim, d\Omega \rangle 
=0$, and thus $\widehat{X}\spa\rest_\scrim$ is tangent to $\scrim$ as wanted. \\ 
The set on $\scrim$ of the points where $\widehat{X}$ vanishes is closed since $\widehat{X}$ is continuous.
To conclude, we wish to prove that $\widehat{X}\spa\rest_\scrim$
cannot vanish on every (nonempty) open set $A\subset \scrim$ (otherwise it vanishes everywhere on $\scrim$, 
but this case is not allowed by definition of $X$).
Assume that there is such $A$ where  $\widehat{X}\spa\rest_A=0$, take $p\in A$ and fix any other point $q\in 
\scrim$, such that there is a  $\hg$-geodesics, $\gamma \subset \scrim$,  joining $p$ and $q$. We assume here 
that $\gamma$ is either a space-like geodesics on $\bS^2$ or a null-like geodesic at constant angular 
variables. We want to prove that $\widehat{X}(q)=0$ when $\widehat{X}\spa\rest_A=0$.\\
If $\widehat{X}\spa\rest_A=0$, all the derivatives $\widehat{\nabla}_a \widehat{X}^b$ vanish, in $A$, when 
$a \neq \Omega$, that is referring to directions tangent to $\scrim$. However, on $\scrim$ it holds 
$\cL_{\widehat{X}} \hg = 0$, by hypotheses. Writing down these equations explicitly, one finds that $\widehat
{X}=0$ on $A$ implies $\widehat{\nabla}_\Omega \widehat{X}^b=0$ if $b\neq \Omega$. However $\widehat{\nabla}_
\Omega X^\Omega\spa\rest_\scrim =0$ holds since both $X^\Omega = X(\Omega)$ and $X(\Omega)/\Omega =X^\Omega/
\Omega$ vanishes on $\scrim$. We have found that, in $A$, $\widehat{F}_{ab}=0$. Notice that $\varphi =0$ in 
$A$, since it is proportional to the limit of $\Omega^{-1}X(\Omega)$ approaching $\scrim$ which vanishes by 
hypotheses. This also entails that $\widehat{K}_a=0$ when $a \neq \Omega$, in $A$, that is $\widehat{K}^a 
\neq 0$ for $a= \ell$ at most, in $A$. Let $k$ denote the value $\widehat{K}(p)$ for the considered field 
$\widehat{X}$ with $\widehat{X}\spa\rest_A=0$. Let us finally focus on the differential equations (\ref{OPEC}
) referred to the mentioned  geodesic $[0,1] \ni t \mapsto \gamma(t)$. We argue that a solution, and thus the
unique solution, for initial data at $p$,
$\widehat{X}(0) =0$, $\widehat{F}_{ab}(0)=0$,  $\widehat{\varphi}(0)=0$, $\widehat{K}(0):= k$ is $\widehat{X}
(t) =0$, $\widehat{F}_{ab}(t)=0$,  $\widehat{\varphi}(t)=0$, $\widehat{K}(t)$, for all $t\in [0,1]$,
where the last function is the unique satisfying $\dot{\gamma}^a \widehat\nabla_a \widehat{K}_b=0$
with $\widehat{K}(0):= k$. To prove it notice that, inserting these functions in (\ref{OPEC}), the equations 
reduce to
\beq 
 \dot{\gamma}^a \widehat{K}_a =0 \:, \quad \dot{\gamma}^a \widehat{K}^{b} - \dot{\gamma}^b \widehat{K}^a =0\:, \quad 
 \dot{\gamma}^a \widehat\nabla_a \widehat{K}_b = 0, \label{OPEC2}
\eeq
The first two equations are certainly fulfilled at $t=0$ by hypotheses, the third one determines $K$ uniquely with the initial condition
$\widehat{K}(0):= k$. However also the first two equations are fulfilled on this solution in view of the fact that they are fulfilled at
$t=0$ and that $\dot{\gamma}^a \widehat{\nabla}_a \dot{\gamma}^b=0$ since we are dealing with a  geodesic. 
We have found that, in particular, $X$ vanishes at $q$ as wanted, since $X(1)=0$. With the same procedure, moving $p$ and $q$ about the original positions,
we find that $X$ vanishes in a open set $A_q$ which enlarges $A$ and it includes $q$. Iterating the 
procedure, we can enlarge $A_q$ in order to include any third point $q'\in \scrim$, joined to $q$ by means of 
a second geodesics, so that  $X$ vanishes at $q'$ too.
In view of the form (\ref{quasih}) of the metric on $\scrim$, for every couple of points $p,q'\in \scrim$, there is always a 
sequence of three consecutive geodesics, of the two above-mentioned types, joining $p$ and $q'$. Therefore $X$
vanishes everywhere on $\scrim$.
\\
(b) In a neighbourhood of $\scrim$, referring to coordinates $\Omega, \ell, \theta,\phi$ one has
$$\widehat{X} = f^\Omega \partial_\Omega+ f^\ell \partial_\ell + f^{\theta} \partial_\theta + 
f^{\phi} \partial_\phi\:.$$
Approaching $\scrim$ (i.e. as $\Omega=0$) one gets (1) $f^\Omega=0$, since $\widehat{X}$ becomes tangent to $\scrim$.
However one also finds (2) $\partial_\Omega f^\Omega\spa \rest_\scrim=0$ as a consequence of $
(f^\Omega - f^\Omega\spa\rest_\scrim)/\Omega =
\Omega^{-1}X(\Omega) \to 0$ approaching $\scrim$.
Since $\widehat{X}\spa\rest_\scrim$ is tangent to the null surface $\scrim$ and it is the limit of a timelike vector,  we also know that, at the points where it does not vanish, it must be  light-like and future directed. Since
$\widehat{X}\spa\rest_\scrim= f^\ell \partial_\ell + f^{\theta} \partial_\theta + 
f^{\phi} \partial_\phi$, the requirement $\hg(\widehat{X},\widehat{X})\spa \rest_\scrim=0$ implies that (3) $f^{\theta}=f^{\phi} =0$ everywhere on $\scrim$,
in view of the Bondi form of the metric on $\scrim$. Therefore (4) $\widehat{X}\spa\rest_\scrim =
f^\ell(0, \ell, \theta, \phi) \partial_\ell$.
 Using Bondi form of the metric again, the requirement $(\cL_{\widehat{X}} \hg)\spa\rest_\scrim = 0$ produces immediately the constraints $\partial_\ell f^\ell\spa\rest_\scrim  =0$ in view of (1),(2), (3), and (4), so that 
$\widehat{X}\spa\rest_\scrim= f(\theta,\phi) \partial_\ell$. Since $\widehat{X}\spa\rest_\scrim$ cannot
vanish in any open set on $\scrim$, $f$ cannot vanish in any open set on $\bS^2$. Since $f$ is smooth and
thus continuous, the set $f^{-1}(0)$ must be closed. Since, with our sign convention for the Bondi metric, both 
$X$ and $\partial_\ell$ are future oriented, $f$ cannot be negative.
$\Box$\\

\noindent {\bf Proof of Proposition \ref{togroup}}. We start from the proofs of (a) and (b). 
If there were a smooth extension of $Y$ to $\overline{M} = M \cup \scrim$ it would be unique by continuity 
and it would satisfy $\cL_{\widehat{Y}} \widehat{g} =0$ up to $\scrim$ by continuity again. Therefore it is 
sufficient to establish the existence of  a smooth extension to $\hM$ to get the most relevant part of (a) 
and (b). The proof is essentially the same as done in the proof of Proposition \ref{X}, concerning the 
existence of the extension of the field $X$. Now, $Y$ is a proper conformal Killing field so that the 
transport equations (\ref{OPEC2}) \cite{Geroch2,Hall}  reduces to 
\beq \dot{\gamma}^a \widehat\nabla_a \widehat{Y}_b = \dot{\gamma}^a \widehat{F}_{ab} \quad\mbox{and} \quad
\dot{\gamma}^a \widehat\nabla_a \widehat{F}_{bc} = \widehat{R}_{bcad}\dot{\gamma}^a 
\widehat{Y}^d\:,\label{OPE}\eeq
The procedure is exactly as that in the proof of Proposition \ref{X} and,
in this way, one obtains a smooth extention $\widehat{Y}$ of $Y$ on $\hM$ and in particular on $\scrim$.
The condition that $\widehat{Y}$ is tangent to $\scrim$
is $\langle \widehat{Y}, d\Omega\rangle =0$ everywhere on $\scrim$. However $g^{sb} \partial_b \Omega = (\partial_\ell)^s$ and
$X \to f\partial_\ell$ approaching $\scrim$, for some nonnegative function  $f \in C^\infty(\bS^2)$, as showed in Proposition \ref{X}. Therefore 
$ \langle \widehat{Y}, d\Omega\rangle f = \lim_{\to \scrim} g(\widehat{Y}, X)$.
If the limit vanishes approaching $\scrim$, $\langle \widehat{Y}, d\Omega\rangle=0$ on the points $(\ell,s)
\in \bR \times \bS^2$ where $f(s)\neq 0$. This happens on an open nonempty set $B\subset \bS^2$.
Therefore $\langle \widehat{Y}, d\Omega\rangle=0$ on $\bR \times B$.
Let $(\ell_0,s_0) \not \in \bR \times B$. Since $\bS^2 \setminus B$ has no interior (see Proposition 
\ref{X}), there is a sequence $ \bR \times B \ni (\ell_0, s_n) \to (\ell_0,s_0)$ as $n\to \infty$. 
Continuity of $(\ell,s) \mapsto \langle \widehat{Y}, d\Omega\rangle(\ell,s)$ implies 
$\langle \widehat{Y}, d\Omega\rangle=0$ in $\bR \times (\bS^2 \setminus B)$ and, thus, everywhere.
Conversely, if  $\widehat{Y}$ is tangent to $\scrim$, then
$\langle \widehat{Y}, d\Omega\rangle =0$ on $\scrim$, and hence $ \lim_{\to \scrim} g(\widehat{Y}, X) = 
\langle \widehat{Y}, d\Omega\rangle f =0$. \\
To conclude, we prove the last statements: (c) and (d).
Since the map $Y \mapsto \widehat{Y}\spa\rest_\scrim$ 
is linear by construction, (d) is a trivial consequence of (c). Let us prove (c).
  If the considered space is made of the zero vector only,
the proof of (c) is trivial. Assume that it is not the case. To prove (c), it is
sufficient to prove that the identity $\widehat{Y}\spa\rest_\scrim =0$ on a set $A\subset \scrim$ which is nonempty and open with respect to 
the topology of $\scrim$, entails $Y=0$ in $M$ (and thus $\widehat{Y}=0$ in $M \cup \scrim$ by continuity). 
Let us show it.
Consider any fixed
 point $p \in M$ and a smooth path $\gamma$ from some $q\in A$ to $p$ (it exists because $M$ is connected and $\scrim= \partial M$). 
 In view of the first order 
 transport equations (\ref{OPE}),
$Y(p)=\widehat{Y}(p) = 0$ when both $\widehat{Y}(q)$ and $\widehat{F}_{ab}(q)$ vanish. Let us show that it is the case.
Suppose that  $\widehat{Y}\spa\rest_\scrim =0$ on $A$ as above. Using coordinates $(\ell,\Omega, \theta, \phi)$
about $\scrim$, one has that $\partial_a \widehat{Y}^b\spa\rest_A=0$ if $a\neq \Omega$. On the other hand, the condition 
$\cL_{\widehat{Y}} \widehat{g}_{ab}=0$ computed on $A$, taking into account $\widehat{Y}\spa\rest_A=0$ and 
$\partial_a \widehat{Y}^b\spa\rest_A=0$ if $a\neq \Omega$, yields
$\partial_\Omega \widehat{Y}^b\spa\rest_A=0$, so that $\widehat{\nabla}_a \widehat{Y}^b\spa\rest_A = \partial_\Omega 
\widehat{Y}^b \spa\rest_A+ 
\widehat{\Gamma}^b_{ac} \widehat{Y}^c\spa\rest_A = 0$. Therefore $F_{ab}\spa\rest_A =0$ and it concludes the proof. 
$\Box$\\

\noindent {\bf Proof of Proposition \ref{defSG}}. (a) If $(s^1,s^2)$ are (local) coordinates of a point $s\in \bS^2$, 
fix $\alpha,\beta \in C^\infty(\bS^2)$ and real constants $r_1,r_2,r_3$. We wish to study  
the integral lines $t \mapsto (\ell(t), s(t))\in \bR \times \bS^2$ of the field 
$Z(\ell, s):= (\alpha(s) \ell  + \beta(s)) \partial_\ell  + \sum_{k=1}^3 r_k S_k^i \partial_{s^i}$
on $\bR\times \bS^2$, with initial condition $(\ell_0,s_0)$.
By construction, the components referred to the sphere do not depend on $\ell$ and thus, the corresponding equations can be
integrated separately. Since  $\sum_{k=1}^3 r_k S_k^i \partial_{s^i}$ is smooth and $\bS^2$ is compact, the integral 
lines $t \mapsto s(t|s_0)$ (here and henceforth $|s_0$ denotes the initial condition at $t=0$) must be smooth and
complete (i.e. defined for $t\in (-\infty,+\infty)$), in view of well-known theorems of differential equations on manifolds. 
Then assume that the smooth function $\bR \ni t \to s(t|s_0)$ is known (computed as above). The remaining differential equation reads
$$\frac{d\ell}{dt} = \alpha(s(t|s_0)) \ell  + \beta(s(t|s_0))\:.$$ 
It can be integrated and the right-hand side is defined for the values of $t$ where the full integral '
converges:
\beq
\ell(t|s_0,\ell_0) = e^{\int^t_0 dt_1 \alpha(s(t_1|s_0))}   \ell_0 + e^{\int^t_0 dt_1 \alpha(s(t_1|s_0))}\int_0^t dt_1 \beta(s(t_1|s_0)) 
e^{-\int_0^{t_1} dt_2 \alpha(s(t_2|s_0))}.  \label{formulazza}
\eeq
It is apparent that the parameter $t$ ranges in the whole real axis due to smoothness of $\bR\ni t \to \alpha(s(t|s_0))$ and
$\bR \ni t\to \beta(s(t|s_0))$, and that $\bR \ni t \mapsto \ell(t|s_0,t_0)$ is smooth as well. 
We have established that the integral lines of 
$Z$ are complete and thus, in view of known theorems, the one-parameter group of diffeomorphisms generated by $Z$ is global.
Since $s=s(t)$ must necessarily describe a rotation of $SO(3)$, about the axis $(r_1,r_2,r_3)/\sqrt{r_1^2+r^2
_2+r^2_3}$ with angle $t\sqrt{r_1^2+r^2_2+r^2_3}$, of the point on $\bS^2$ initially individuated by $s_0$ 
and, taking (\ref{formulazza}) into account, it is evident that each diffeomorphism 
 $$\bR  \times \bS^2 \ni (\ell_0,s_0) \mapsto (\ell(t|s_0,t_0),s(t|s_0)) \in \bR \times \bS^2\:,$$ 
for every fixed $t\in \bR$, has the form (\ref{group}) and, thus, it belongs to $SG_\scrim$.\\
(b) A fixed $(a,b,R)\in SG_\scrim$ can be decomposed as
$$(R, a,b) = (I, a\circ R^{-1}, b\circ R^{-1})\: (R,0,0)\:.$$
Looking at (\ref{formulazza}), $(R,0,0)$ is an element of the one-parameter group 
generated by $\sum_{k=1}^3 n_k S_k$, where $(n_1,n_2,n_3)$ are the Cartesian
components of the rotation axis of $R$; conversely the transformation $(I, a\circ R^{-1}, b\circ R^{-1})$ can 
be written as  $\exp\{1Z\}$ where 
$Z= \ell a\left(R^{-1}(s)\right) \partial_\ell + b\left(R^{-1}(s)\right) \partial_\ell$. $\Box$\\

\noindent {\bf Proof of Theorem \ref{theorem2}}. Consider the {\em local} one-parameter group of 
diffeomorphisms generated by $\widehat{Y}$ in a sufficiently small neighbourhood (in $\hM$) of a point  $q\in 
\scrim$ and for $t\in (-\epsilon,\epsilon)$ with $\epsilon>0$ sufficiently small.
In local coordinates over $\scrim$, $(\ell,s^1,s^2) \in (a,b)\times A$, such a set of transformations can be represented by
\beq
\ell \to \ell_t := f(\ell, s_1, s_2, t)\:, \quad (s^1,s^2) \to (s^1_t,s^2_t) := g(\ell, s^1,s^2, t)
\quad \mbox{with $(\ell,s^1,s^2) \in (a,b)\times A$.}\label{diffgen2}
\eeq
Using the same argument as the one used to characterise the group $SG_\scrim$ (after Definition \ref{defpreservscri}), one 
finds that
it must be $g(\ell, s_1,s_2,t) = R_t(s)$ for all $\ell, s$
and $f(\ell, s_1,s_2,t)= c(s_1,s_2,t) \ell + b(s_1,s_2,t)$, for all $\ell, s$, for some $R_t\in O(3)$ 
depending on $t$ smoothly, and where $c,b$ are jointly smooth real functions. The requirement, that 
$t\mapsto R_t$ is a (local) one-parameter subgroup of $SO(3)$, implies that $\frac{dR_t}{dt}|_{t=0} = \sum_{k=1}^3 r_k 
S_k(s_1,s_2)$. Similarly $\frac{df_t}{dt}|_{t=0} = \frac{\partial c(s_1,s_2,t)}{\partial t}|_{t=0} \ell + 
\frac{\partial b(s_1,s_2,t)}{\partial t}|_{t=0}$. We have found that, in local coordinates
$$\widehat{Y}\spa \rest_\scrim = \sum_{k=1}^3 r_k S_k(s_1,s_2) + 
\frac{\partial c(s_1,s_2,t)}{\partial t}|_{t=0} \ell \partial_\ell + \frac{\partial b(s_1,s_2,t)}{\partial t}|_{t=0}\partial_\ell\:,$$
and thus, about $q$, $\widehat{Y}\spa \rest_\scrim$ takes the form of the vectors in $\gg_\scrim$. However,
since it holds true in a neighbourhood of each point on $\scrim$, we have that $\widehat{Y}\spa \rest_\scrim \in \gg_\scrim$.\\
To conclude, (b) is an immediate consequence of (a) and of the last part of (a) in Proof of Proposition 
\ref{defSG}. $\Box$\\

\noindent {\bf Proof of Proposition \ref{positiveenergy}}. 
  Since $\widetilde{Y}\in \gg_\scrim$, in principle  it has the form
$$\widetilde{Y}(\ell,s) = \sum_{i=1}^3 c_i S_i(s) + (f(s)+ \ell g(s))\partial_\ell\:.$$
Since $\hg(Y,Y)<0$ about $\scrim$ and its limit toward $\scrim$, namely $\widetilde{Y}$, is tangent to 
$\scrim$ it must satisfy $\hg(\widetilde{Y},\widetilde{Y})=0$ by continuity 
(no timelike tangent vectors can be tangent to a null surface).
Using the form (\ref{quasih}) of $\hg$ one see that it must be:
$\sum_{i=1}^3 c_i S_i(s)=0$ on $\scrim$.
Using the explicit form of $S_1,S_2,S_3$ referring to the base $\partial_\phi,\partial_\theta$ of $T\bS^2$, one sees that this 
is equivalent to claim that, everywhere on the sphere,
$$(c_1 \sin \phi -c_2 \cos \phi) =0\:,\quad c_1 \cot \theta \cos \phi + c_2 \cot\theta \sin \phi + c_3=0$$
As a consequence $c_1= c_2= c_3 =0$. Therefore, everywhere on $\scrim$
$$\widetilde{Y}= (f(s)+ \ell g(s))\partial_\ell\:,$$
for some functions $f,g \in C^\infty(\bS^2)$.
$\widetilde{Y}$ is the limit of a causal future-directed vector. Therefore, it has either to vanish or to be 
directed as $\partial_\ell$ at every point of $\scrim$. Since $\ell g(s)$ may take every arbitrarily large, 
positive or negative, value (notice that $g$ is bounded, it being smooth on a compact set), it must be $g(s)=
0$ and $f(s)\geq 0$.
$\Box$\\

\noindent {\bf Proof of Theorem \ref{theoremlambdaM}}. As before, from now on, $(\cF_+(\cH),\Pi,\Upsilon)$ is the GNS triple of $\lambda$.
First of all we notice that $\lambda_M$ is in fact a well-defined state on $\cW(M)$ since $\imath$ is a 
$*$-homomorphism. $\lambda_M$ is quasifree associated with a real scalar product $\mu_M :\cS(M) \times \cS(M) \to \bR$ defined as $\mu_M(\varphi,\varphi') := \mu(\gamma\Gamma\varphi,\gamma\Gamma\varphi')$. From this fact, it follows
that the  GNS triple of $\lambda_M$ can be constructed as $(\cF_+(\cH_M),\Pi\spa\rest_{A_M},\Upsilon)$,
where $A_M\subset \cW(\scrim)$ is the sub $C^*$-algebra isomorphic to $\cW(M)$ in view of Theorem 
\ref{theoremimath}, $\cH_M$ is the Hilbert subspace of $\cH$ given by the closure of the space of complex
linear combinations of $K_{\mu}(\Gamma(\varphi))$, for every $\varphi\in \cS(M)$ and, thus, $\cF_+(\cH_M)$
is a Fock subspace of $\cF_+(\cH)$. In particular, the canonical $\bR$-linear map $K_{\mu_{M}} :\cS(M) \to \cH_M$ is nothing but $K_{\mu_{M}} = K_{\mu} \circ \gamma\Gamma$.\\
(a) By construction,  using the definition of $\lambda_M$, 
taking advantage of (\ref{central}) as well as of the invariance property of $\lambda$ under the action of 
$SG_\scrim$,
if $a\in \cW(M)$, one has
$$\lambda_M\left(\beta_t^{(Y)}(a)\right) = \lambda\left(\imath\left(\beta_t^{(Y)}(a)\right)\right) = 
\lambda\left(\alpha_t^{(\widetilde{Y})}\imath(a)\right)
= \lambda\left(\imath(a)\right) = \lambda_M(a)\:.$$
This proves the first part of (a). To conclude the proof of (a), let $V^{(\widetilde{Y})}_t :\cH \to \cH$ the
one-parameter group of unitaries that implements $\alpha^{(\widetilde{Y})}_t$ in the one-particle space $\cH$
for $\lambda$.
  From $K_{\mu_{M}} = K_{\mu} \circ \gamma\Gamma$,  (\ref{central}) and the construction of $V$ one has:
$$V_t^{(\widetilde{Y})} K_{\mu_M} \varphi = V_t^{(\widetilde{Y})} K_{\mu} \gamma\Gamma (\varphi)
= K_\mu\left( \gamma\Gamma\left( \varphi \circ \exp\{-t {Y}\}\right)\right)= 
K_{\mu_{M}}\left(  \varphi \circ \exp\{-t {Y}\}\right)\:.$$
We have found that, for every $\varphi \in \cS(M)$, 
$V_t^{(\widetilde{Y})} K_{\mu_M} \varphi= K_{\mu_{M}}\left(  \varphi \circ \exp\{-t {Y}\}\right)\:,$
hence $V_t^{(\widetilde{Y})}$ leaves the one particle space of
$\lambda_M$,  $\cH_M$, invariant and $V_t^{(\widetilde{Y})}\spa\rest_{\cH_M}$ implements $\beta_t^{(Y
)}$ in $\cH_M$. As a consequence of the structure of the GNS triple of $\lambda_M$, if $U_t^{(\widetilde{Y})
}$ implements $\beta_t^{(\widetilde{Y})}$ unitarily in $\gH= \cF_+(\cH)$ leaving $\Upsilon$ invariant, it 
leaves also invariant the structure of the GNS-Fock space of $\lambda_M$ and, therein, $U_t^{(\widetilde{Y})}
\spa\rest_{\cF_+(\cH_M)}$ implements $\alpha_t^{(Y)}$ unitarily in $\gH_M= \cF_+(\cH_M)$ leaving the cyclic
vector invariant. In other words 
$$U_t^{({Y})} = U_t^{(\widetilde{Y})}\spa\rest_{\cF_+(\cH_M)}\:.$$
Notice that $\bR \ni \mapsto U_t^{(\widetilde{Y})}\spa\rest_{\cF_+(\cH_M)}$ is strongly continuous 
since $\bR \ni \mapsto U_t^{(\widetilde{Y})}$ is such.
Moreover the self-adjoint generator of $U_t^{(\widetilde{Y})}\spa\rest_{\cF_+(\cH_M)}$ is obtained by 
restricting that of $U_t^{(\widetilde{Y})}\spa\rest_{\cF_+(\cH_M)}$ to $\cF_+(\cH_M)$. If the former 
generator is positive, the latter has to be so. In the considered case, the former is positive since $Y$ is 
timelike and future directed 
and thus we can apply (a) of Proposition \ref{PP}. The same argument shows that the self-adjoint generator of 
$V_t^{(\widetilde{Y})}\spa\rest_{\cH_M}$ has no zero modes if $V_t^{(\widetilde{Y})}\spa\rest_{\cH_M}$ has 
no zero modes. This last fact happens if $\widetilde{Y}$ vanishes  on a zero-measure subset of $\scrim$ due 
to (b) of Proposition \ref{PP}. $\Box$

\noindent {\bf Proof of Theorem \ref{Nicola}}. (a) Consider a wavefunction $\ph \in \cS(M)$. It satisfies $\ph=Ef$
where $E: C^\infty_0(M) \to \cS(M)$ is the causal propagator and $f$ is some real smooth and compactly 
supported function in $M$. Since the maximally extended de Sitter spacetime $M'$ is globally hyperbolic and 
$M\subset M'$, -- so that  $C^\infty_0(M)\subset C^\infty_0(M')$ -- one can focus on the wavefunction
$\ph' := E'f$, where $E'$ is the causal propagator in  $M'$. By construction $\ph'\spa\rest_M = \ph$, so that 
$\ph'$ is a smooth extension of $\ph$. Since $\scrim \subset M'$, all that implies that $\ph$ extends to 
$\scrim$ smoothly (and uniquely) and this extension is $\lim_{\to \scrim} \ph = \ph'\spa\rest_\scrim$. In 
this way, an $\bR$-linear map $\Gamma: \cS(M) \ni \ph \to \ph'\spa\rest_\scrim \in C^\infty_0(\scrim)$
is defined. To conclude (a), it is enough to prove both that $Ran \Gamma \subset \sS(\scrim)$ and that 
$\Gamma$ preserves the symplectic forms. Let us prove them. Bearing in mind the previously discussed
behaviour of $H^{(2)}_{\nu}(z)$ for large $z$  (with $|arg z|\leq\pi - \epsilon$), making use of (\ref{Phik}) 
and (\ref{chi}), the identity (\ref{FFinv}) can be recast as
\beq
\ph(\tau,\bx) =\frac{e^{-i\frac{\pi}{4}} }{\gamma 4 \pi^{3/2}}  \int_{\bS^2} \sp \epsilon_{\bS^2}(\theta,\phi)\int_0^{+\infty} \sp\sp dk  k
 e^{i(k r \cos \lambda_\bx(\theta,\phi) - k\tau)}\left[\tau + O\left(\frac{1}{k}\right)\right]
\sqrt{k}\widetilde{\ph}(k,\theta,\phi) +  c.c.\:,\label{FFinv2}
\eeq 
where $\lambda_\bx(\theta,\phi) \in [0,\pi]$ is the angle between $\bx$ and $\bk$. The iterated integrations make 
sense and can be interchanged (via Fubini-Tonelli theorem) since both $\sqrt{k}\widetilde{\ph}(k,\theta,
\phi)$ and  $(\sqrt{k}\widetilde{\ph}(k,\theta,\phi)$ are integrable in the measure $d\bk$. They are smooth 
everywhere but $\bk=0$, they vanish very fast at large $|\bk|$ and, for $\bk=0$, $\widetilde{\ph}\propto 
1/|\bk|^{-Re |\nu|}$ if $m^2 + \xi R > 0$ for $\nu$. Now, calling $\tau = (u+v)/2$ and $r= (u-v)/2$, $\scrim$
arises as the limit $v\to -\infty$. The contribution due to the factor of $O\left(\frac{1}
{k}\right)$ vanishes due to the Riemann-Lebesgue lemma:
\beq
 \left(\Gamma \ph\right)(u,\theta_\bx,\phi_\bx)= \lim_{s\to +\infty}
\frac{e^{-i\frac{\pi}{4}}}{\gamma 4 \pi^{3/2}} \int_0^{+\infty} \sp\sp\sp dk \int_{\bS^2} \sp \epsilon_{\bS^2}(\theta,\phi)  
\frac{ks}{2} e^{i\frac{ks}{2}[\cos \lambda_\bx(\theta,\phi)+ 1]} e^{-iuk }
\sqrt{k}\widetilde{\ph}(k,\theta,\phi) +  c.c.\nonumber 
\eeq
That limit can be computed using integration by parts exactly as in the appendix A2 of \cite{DMP}. In detail,
one rotates the axes so that the axis $z$ coincides with $\bx$ and, thinking of $\widetilde{\ph}$
as a function of $k,c,\phi$ where $c:= \cos \theta\in [-1,1]$, one  re-arranges the expression above as 
\beq
\left(\Gamma \ph\right)(u,\theta_\bx,\phi_\bx)= \lim_{s\to +\infty}
\frac{-i e^{-i\frac{\pi}{4}}}{\gamma 4 \pi^{3/2}} \int_0^{+\infty} \sp\sp\sp dk \int_{0}^{2\pi} 
\sp\sp\sp d\phi  \int_{-1}^1 
\sp\sp dc
\frac{\partial}{\partial c}\left( e^{i\frac{ks}{2}[c + 1]} \right) e^{-iuk }
\sqrt{k}\widetilde{\ph}(k,c,\phi) +  c.c.\nonumber 
\eeq
where $\theta_\bx = 0$ in our case. 
The right-hand side can be expanded using integration by parts and only the contribution for $c=-1$
(that is $\theta = -\pi$, i.e. $\bk/|\bk| = -\bx/|\bx|$)
survives, the others vanish as $s\to +\infty$, due to Riemann-Lebesgue's lemma (interchanging various integrations using Fubini-Tonelli theorem and finally taking advantage of dominate convergence theorem).
The integration over $\phi$ produces a trivial factor $2\pi$ since the dependence from $\phi$ 
of the involved functions disappears as $\theta=0,\pi$.
 The final result reads, using the initial generic choice for the axes $x,y,z$:
 \beq
 \left(\Gamma \ph\right)(u,\theta_\bx,\phi_\bx)= \frac{i2\pi e^{-i\frac{\pi}{4}}}{\gamma 4 \pi^{3/2}}  \int_0^{+\infty} \sp\sp dk  \: e^{-iuk }\:\:
\sqrt{k}\widetilde{\ph}(k,\eta(\theta_\bx,\phi_\bx)) +  c.c.\:,\nonumber 
\eeq
 $\eta : \bS^2 \to \bS^2$ denoting the parity inversion $\bS^2 \ni
{\bf n} \mapsto -{\bf n}\in \bS^2$.
Dropping the index $\bx$, and viewing $\theta,\phi$ as the standard coordinates on $\scrim$, the 
obtained result can be re-written as
\beq
 \left(\gamma\Gamma \ph\right)(\ell,\theta,\phi)= i 
\frac{e^{-i\frac{\pi}{4}}}{(-\gamma)} \int_0^{+\infty} \sp\sp dk  \: \frac{e^{-i\ell k }}{\sqrt{2\pi}}\:\:
\sqrt{\frac{k}{2(-\gamma )}}\widetilde{\ph}\left( \frac{k}{(-\gamma)},\eta(\theta,\phi)\right) +  c.c.\:.\label{final1} 
\eeq
where we have passed to the standard Bondi coordinates on $\scrim$, i.e. $\ell,\theta,\phi$ with $u = -\gamma \ell$.
In our hypotheses on $\ph$ and $\nu$, most notably $m^2+\xi R >\frac{5}{48}R$, the functions
$\sqrt{\frac{k}{2}}\widetilde{\ph}(k,\eta(\theta,\phi))$ and $k\sqrt{\frac{k}{2}}\widetilde{\ph}(k,\eta(
\theta,\phi))$ belong also to $L^2(\bR^+\times \bS^2; dk\wedge \epsilon_{\bS^2}(\theta,\phi))$.  This implies 
that both the functions $\Gamma\ph, \partial_\ell\Gamma\ph$ belong to $L^2(\bR\times\bS^2; d\ell\wedge
\epsilon_{\bS^2})$. In this way we have found that $Ran\Gamma\subset\sS(\scrim)$. Actually we have obtained 
much more: by means of both (\ref{Aadded}) and the Fourier transformed expression of $\sigma$, (\ref{final1}) 
implies that
\begin{gather*}
\sigma(\gamma \Gamma \ph, \gamma \Gamma \ph') = -2 Im \left\{(-\gamma)^{-2}\int_{\bR^+\times \bS^2} \sp\sp\sp dk \wedge \epsilon_{\bS^2} 2k 
\frac{k}{2(-\gamma )}\overline{\widetilde{\ph}\left( \frac{k}{(-\gamma)},\eta(\theta,\phi)\right)}
\widetilde{\ph'}\left( \frac{k}{(-\gamma)},\eta(\theta,\phi)\right)
 \right\}\\
 = -2 Im \left\{\int_{\bR^+\times \bS^2} \sp\sp\sp k^2 dk \wedge \epsilon_{\bS^2} 
\overline{\widetilde{\ph}( k,\theta,\phi)}
\widetilde{\ph'}( k,\theta,\phi)\right\}= 
-2 Im \left\{\int_{\bR^3} d\bk
\overline{\widetilde{\ph}(\bk)}
\widetilde{\ph'}(\bk)\right\} = \sigma_M( \ph, \ph')\:,
\end{gather*}
where in the last step we exploited (\ref{ImsigmaM}). Hence $\gamma \Gamma$ preserves the symplectic form as 
requested.\\
(b) Exactly as in the last step of the proof of (a), since the functions
$\sqrt{\frac{k}{2}}\widetilde{\ph}(k,\eta(\theta,\phi))$ and $k\sqrt{\frac{k}{2}}\widetilde{\ph}(k,\eta(
\theta,\phi))$ are also in  $L^2(\bR^+\times \bS^2; dk\wedge \epsilon_{\bS^2}(\theta,\phi))$,  
(\ref{scalar}) and (\ref{final1}) imply:
\begin{gather*}
\mu(K_\lambda \gamma\Gamma \ph, K_\lambda \gamma\Gamma \ph)  = (-\gamma)^{-2}\int_{\bR^+\times \bS^2} \sp\sp\sp dk \wedge \epsilon_{\bS^2} 2k 
\frac{k}{2(-\gamma )}\overline{\widetilde{\ph}\left( \frac{k}{(-\gamma)},\eta(\theta,\phi)\right)}
\widetilde{\ph}\left( \frac{k}{(-\gamma)},\eta(\theta,\phi)\right)\\
= \int_{\bR^+\times \bS^2} \sp\sp\sp k^2 dk \wedge \epsilon_{\bS^2} 
\overline{\widetilde{\ph}( k,\theta,\phi)}
\widetilde{\ph}( k,\theta,\phi)= 
\int_{\bR^3} d\bk
\overline{\widetilde{\ph}(\bk)}
\widetilde{\ph}(\bk)
\end{gather*}
Therefore, for every $\ph\in \cS(M)$, in view of (\ref{OmegaE}),
$$\lambda_M(W_M(\ph)) := \lambda(W(\gamma\Gamma \ph)) = e^{-\mu(K_\lambda \gamma\Gamma \ph, K_\lambda \gamma\Gamma \ph)/2}
= e^{-\frac{1}{2}  \int_{\bR^3}  \overline{\widetilde{\ph}(\bk)}\widetilde{\ph}(\bk) \: d\bk } =
\omega_E(W_M(\ph))\:, $$
and this concludes the proof. $\Box$\\

\noindent {\bf Proof of Theorem \ref{Nicola2}}. Here, we exploit the same notation, {\it i.e.} $\bx,\bk$, as in the proof of Theorem 
\ref{Nicola}.
In particular  $\nu := \sqrt{\frac{9}{4}-(m^2 \gamma^2 + 12\xi)}$, so that $\nu\geq 0$ when 
$\frac{9}{4}-(m^2 \gamma^2 + 12\xi)\geq 0$ in the following. However the sign of $\nu$ could be fixed arbitrarily
 (and this applies for imaginary 
$\nu$, in particular), since the functions we shall employ are invariant under $\nu \to -\nu$.\\
 As a first step, we notice that if $\ph\in \cS(M)$, it extends to $\scrim$ smoothly so that 
$\Gamma \ph := \lim_{\to \scrim}\ph \in C^\infty(\scrim)$ does exist. This is because, as found in the section
\ref{FRWsection},
the spacetime $(M,g)$ extends to a larger spacetime equipped with a metric $\hg$ 
obtained by multiplying the metric of the closed static Einstein universe with a strictly positive smooth 
factor. Since closed static Einstein universe is globally hyperbolic and global hyperbolicity does not depend on  nonsingular 
conformal rescaling of the metric, $(M,g)$ itself is included in a globally hyperbolic spacetime. With the same argument used 
for de Sitter spacetime in the proof of Theorem \ref{Nicola}, one has that every $\ph\in \cS(M)$ extends to $\scrim$ smoothly.
We have now to show that $Ran \Gamma \subset \sS(\scrim)$ and that $\Gamma$ preserves the symplectic forms.\\ 
First of all, analogously to what done in the de Sitter case,  we determine a class of modes $\Psi_\bk(\tau,
\bx)$ that will be useful in decomposing the solutions of Klein-Gordon equation in order to take the limit of 
wavefunctions towards $\scrim$. 
\beq \label{rhok}
\Psi_\bk(\tau,\bx) := \frac{e^{i\bk\cdot\bx}}{(2\pi)^{3/2}} \frac{\rho_\bk(\tau)}{a(\tau)}\:,
\eeq 
where, taking the exponential factor into account, the Klein-Gordon equation reduces to the following equation 
for the functions $(-\infty,0)\ni \tau \mapsto \psi_\bk(\tau)$,  
\begin{eqnarray}
&& \frac{d^2}{d\tau^2}\rho_\bk(\tau)  + (V_0(\bk,\tau)  + V(\tau)) \rho_\bk(\tau) =0,  \nonumber \\ 
&&\mbox{with}\quad V_0(\bk,\tau):= 
k^2 + \left(\frac{\gamma}{\tau}\right)^2\left[m^2 + \left(\xi-\frac{2}{\gamma^2} \right) \right], 
 \quad V(\tau)= O(1/\tau^3)\:. \label{rhokeq}
\end{eqnarray}
Comparing with Klein-Gordon equation, one sees that  $V_0(\bk,\tau)  + V(\tau) = k^2 + a(\tau)^2[m^2 + (\xi-1/6)R(\tau)]$
where $V_0$ is nothing but the the contribution of pure de Sitter metric and $V$ is a perturbation.
If we dropped the perturbation $ V(\tau)$, 
the functions $\rho_\bk$ would reduce to the functions $\chi_\bk$ and 
the modes $\Psi_\bk$ would reduce to the modes $\Phi_k$ used to construct $\omega_E$ beforehand. Notice that 
the curvature of the spacetime does not coincide with $12/\gamma^2$ as in de Sitter spacetime, but it reads 
$R(\tau)= 12/\gamma^2 + O(1/\tau)$ and $a(\tau)=\ga/\tau +O(1/\tau^2)$. It follows that the added potential 
$V(\tau) = O(1/\tau^3)$ above. A formal solution of (\ref{rhokeq}) is obtained in terms of the series:
\begin{gather}
\rho_\bk(\tau) = \chi_\bk(\tau)\notag\\
+ (-1)^{n} \spa\sum_{n=1}^{+\infty}\spa \int_{-\infty}^{\tau}\sp\sp\sp dt_1
\int_{-\infty}^{t_1}\sp\sp\sp dt_2 \cdots  \int_{-\infty}^{t_{n-1}}\sp\sp\sp \sp dt_n
S_\bk(\tau,t_1)S_\bk(t_1,t_2)\cdots S_\bk(t_{n-1},t_n) V(t_1)V(t_2)\cdots V(t_n)\chi_\bk(t_n),\label{serie}
\end{gather}
where 
\beq
S_\bk(t,t'):= -i \left(\overline{\chi_\bk(t)}\chi_\bk(t')- \overline{\chi_\bk(t')}\chi_\bk(t)\right)\:,\quad 
t,t'\in (-\infty,0)\:, \label{Sk}
\eeq
satisfying, in view of antisymmetry  and (\ref{norm}),
\beq S_\bk(t,t) =0 \quad\mbox{and}\quad
\left.\frac{\partial}{\partial t}S_\bk(t,t')\right|_{t'=t} = 1\label{norm2}\:.
\eeq
By direct inspection and making use of (\ref{norm2}), one sees that the right-hand side of (\ref{serie}) 
defines a solution of 
(\ref{rhokeq}) if one is allowed to interchange the $\tau$-derivative operator -- up to the second order -- 
with the sign of sum. This is always possible  when the series itself and the series of the derivatives of 
first and second order converge $\tau$-uniformly in a neighbourhood of every fixed $\tau \in (-\infty,0)$. 
Actually the locally $\tau$-uniform convergence of the series of derivatives of second order directly follows
from the uniform convergence of those of zero and first order, when one refers to the solutions $\chi_\bk$ 
and the solutions $S_\bk$. Using the expression (\ref{chi}) of the modes $\chi_\bk$, expanding $H_\nu^{(2)}$ 
in terms of Bessel functions $J_{\pm\nu}$ \cite{Grad} and, finally, exploiting standard integral 
representations valid for $Re \nu >-1/2$ (formula 5 in 8.411 in \cite{Grad}) of $J_{\nu}$, one achieves the 
following bounds for $Re \nu< 1/2$  (that is  $m^2 \gamma^2 + 12 \xi>2$), for $\tau<-1$, and for some 
constant $C_\nu\geq 0$
 \beq
 \left.\begin{array}{l}
 |\chi_\bk(\tau)| \leq C_\nu {(-\tau)}^{Re\nu+1/2}\at k^{Re\nu}+k^{-Re\nu}\right) \qquad
 \left|\frac{d \chi_\bk(\tau)}{d\tau}\right| \leq 
 C_\nu {(-\tau)}^{Re\nu+1/2}\at k^{Re\nu}+k^{-Re\nu} \right) (1+k),\label{stime}
\end{array}\right.\eeq

 where $k=|\bk|$.
Furthermore, for the same reasons it is possible to obtain the
following (non optimal) $\bk$-uniform 
bound  for $Re \nu< 1/2$, for $t_2\leq t_1<-1$, and for some other constant $C'_\nu\geq 0$
 \beq
 |S_\bk(t_1,t_2)| \leq C'_\nu {(t_1 t_2)}^{Re\nu +1/2}\:.
 \label{stime1}
 \eeq
Now fix any  $T<-1$ and consider $\tau \in (-\infty,T]$, so that $|V(\tau)| \leq K_T/(-\tau)^3$,
 for some constant $K_T\geq 0$.
  From (\ref{stime}), one sees with a few of trivial computations, that the series in the right-hand side of (\ref{serie})
 and that of the $\tau$-derivatives
 are $\tau$-uniformly dominated, respectively, by 
\beq\label{dominance}
\at k^{Re\nu}+k^{-Re\nu} \ct\; S_{\nu,T}\:,\qquad 
 \at k^{Re\nu}+k^{-Re\nu} \ct (1+k)\; S_{\nu,T}\:,
\eeq
 where $S_{\nu,T}$ is the following convergent series of positive constants
\beq
S_{\nu,T}:=C_\nu \sum_{n=1}^{+\infty} \at\frac{2C'_\nu K_T}{1-2Re\nu}\ct^n \frac{1}{n!} \frac{1}{((-T)^{1-2Re\nu})^{n-1/2}}\:.
\eeq
Summarising, we can conclude that (\ref{serie}) defines a solution of (\ref{rhokeq}) and that, the same
equation entails the solution to be smooth.
As a straightforward consequence we also have the following $\tau$-uniform bound valid on $(-\infty,T]$
\beq
 \left| \rho_\bk(\tau)- \chi_\bk(\tau) \right| \leq \at k^{Re\nu}+k^{-Re\nu} \ct S_{\nu,T}\:, \qquad
 \left| \frac{d\rho_\bk(\tau)}{d\tau}- \frac{d\chi_\bk(\tau)}{d\tau} \right| \leq 2\at k^{Re\nu}+k^{-Re\nu}
  \ct (1+k) S_{\nu,T}.
\eeq
This implies that, at fixed $\tau$, the measurable (since limit of measurable functions) 
functions $\bR^3 \ni \bk \mapsto \rho_\bk(\tau)$  
and $\bR^3 \ni \bk \mapsto \frac{d\rho_\bk(\tau)}{d\tau}$ do not grow, for large $|\bk|$,  fast  than  
$|\bk|^{Re\nu}$ and $|\bk|^{1+Re\nu}$ respectively. Moreover,  
their divergence at $\bk=0$ cannot be worse than that of  $\bR^3 \ni \bk \mapsto \chi_\bk(\tau)$  
and $\bR^3 \ni \bk \mapsto \frac{d\chi_\bk(\tau)}{d\tau}$, that is $k^{-|Re \nu|}$.\\
Finally, notice that each term in the series in the right-hand side of (\ref{serie}) and in the analogy for 
$d\rho_\bk/d\tau$ vanishes as $\tau \to -\infty$ by construction. In view of the fact that, $\tau$-uniformly, 
the series in (\ref{dominance}) dominates both the series in the right-hand side of (\ref{serie}) and 
the series of $\tau$-derivatives, we are allowed to interchange the operations of limit with that of sum, 
obtaining 
\beq
\lim_{\tau\to -\infty}\left( \rho_\bk(\tau)- \chi_\bk(\tau) \right) =0 \quad \mbox{and}\quad
\lim_{\tau\to -\infty}\left( \frac{d\rho_\bk(\tau)}{d\tau}- \frac{d\chi_\bk(\tau)}{d\tau} \right) =0 \:. \label{limiti}
  \eeq 
  This result has a first important consequence. Using equation (\ref{rhokeq}), one sees that the function 
  $\tau \mapsto \frac{d \overline{\rho_\bk(\tau)}}{d\tau}  \rho_\bk(\tau)- \overline{\rho_\bk(\tau)}
\frac{d \rho_\bk(\tau)}{d\tau}$ is actually a constant. The value of this constant can be computed by taking the limit as
 $\tau \to -\infty$, making use of (\ref{norm}), (\ref{limiti}) and taking into account the fact that, for 
 $\bk$ fixed, $\frac{d\rho_\bk(\tau)}{d\tau}$ and $\rho_\bk(\tau)$ are bounded on $(-\infty,T]$ (notice that 
these functions have no limit for $\tau \to -\infty$), as one can show employing the asymptotic behaviour of 
$H^{(2)}_\nu(z)$ for large values of the argument $z$. In this way one finds
 \beq
 \frac{d \overline{\rho_\bk(\tau)}}{d\tau}  \rho_\bk(\tau)- \overline{\rho_\bk(\tau)}
\frac{d \rho_\bk(\tau)}{d\tau} = i \label{norm3}\:.
 \eeq
 Now, to analyse the behaviour of $\Gamma\ph$, we can follow  the same way as that followed in de Sitter space.
  Take any (real by definition) $\ph \in \cS(M)$ and fix a Cauchy surface $\Sigma_{\tau}$ in $(M,g)$ 
  individuated by the points in $M$ with the fixed value of $\tau$; eventually define
\beq
\widetilde{\ph}(\bk) := -i \int_{\bR^3} \left[ \frac{\partial \overline{\Psi_\bk(\tau,\bx)}}{\partial\tau}\ph(\tau,\bx)  - 
 \overline{\Psi_\bk(\tau,\bx)} \frac{\partial \ph(\tau, \bx)}{\partial\tau} \right] a(\tau)^2 d\bx \:. \label{FFFF}
\eeq
The right-hand side of (\ref{FFFF}) does not depend on the choice of $\tau$, as it follows from 
direct inspection, exploiting (\ref{rhokeq}). 
Remembering that $\ph\in \cS(M)$, so that its Cauchy data are real, smooth and compactly supported, 
we have  that their Fourier transform are of Schwartz class. Afterwards, exploiting the fact that 
both the measurable functions $\bR^3 \ni \bk \mapsto \rho_\bk(\tau)$  
and $\bR^3 \ni \bk \mapsto \frac{d\rho_\bk(\tau)}{d\tau}$ grows at most as a polynomial with degree two for large $|\bk|$, and that their divergence at $\bk=0$ is at most of order $k^{-|Re \nu|}$ with $Re \nu<1/2$, we find that  
 $\widetilde{\ph} \in C^{\infty}(\bR^3\setminus\{0\})$ and it vanishes for $|\bk|\to \infty$ faster than every power $|\bk|^{-n}$, $n=1,2,\ldots$. In particular
$\widetilde{\ph}\in L^2(\bR^3;d\bk)\cap L^1(\bR^3;d\bk)$.
Once one knows $\widetilde{\ph}$ by (\ref{FFFF}), the associated $\ph$ can be constructed out of a
decomposition in terms of modes $\Psi_\bk$:
\beq \ph(\tau,\bx) = \int_{\bR^3} \left[ \Psi_{\bk}(\tau,\bx) \widetilde{\ph}(\bk)
+  \overline{\Psi_{\bk}(\tau,\bx)} \overline{\widetilde{\ph}(\bk)}
\right] \: d\bk\:. \label{FFFFinv}
\eeq 
This is a trivial consequence of (\ref{FFFF}), (\ref{rhok}),  (\ref{norm3}), 
and of the standard properties for the Fourier transform of smooth compactly supported 
functions on $\bR^3$. Eventually, per direct computation, one verifies that,
   if $\ph_1,\ph_2 \in \cS(M)$,
   \beq
  -2 Im \left\{ \int_{\bR^3} \overline{\widetilde{\ph}_1}(\bk) \widetilde{\ph}_2  (\bk) d\bk\right\}
  = \int_{\bR^3} (\ph_2 \partial_\tau \ph_1 -  \ph_1 \partial_\tau \ph_2 ) \: a^2(\tau) d\bx =: \sigma_M(\ph_1,\ph_2) \:. 
  \label{ImsigmaM2}
   \eeq
We are now in position to draw some conclusions. Indeed, if $\ph\in \cS(M)$, $p\in \scrim$ and $(\tau_q,
\bx_q)$ are the coordinates of $q\in M$, we can write down
\beq\left(\Gamma\ph\right)(p) = \lim_{q\to p} \int_{\bR^3} \sp\sp d\bk\: \frac{e^{i\bk\cdot\bx_q}}{(2\pi)^{3/2}}
\left(\rho_\bk(\tau_q)-
\chi_\bk(\tau_q)\right) \widetilde{\ph}(\bk)
+ \lim_{q\to p} \int_{\bR^3} \sp\sp d\bk\: \frac{e^{i\bk\cdot\bx_q}}{(2\pi)^{3/2}}
\chi_\bk(\tau) \widetilde{\ph}(\bk)\:\: + c.c.
 \label{2limits}\eeq
As $q\to p\in \scrim$, $\tau_q\to -\infty$ so that $\left(\rho_\bk(\tau_q)-
\chi_\bk(\tau_q)\right)\to 0$ due to (\ref{limiti}). Moreover, since (\ref{dominance}) is valid, we have the $\tau$-uniform bound
$$\left| \frac{e^{i\bk\cdot\bx}}{(2\pi)^{3/2}}
\left(\rho_\bk(\tau)-
\chi_\bk(\tau)\right) \widetilde{\ph}(\bk)\right|  
 \leq \frac{S_{\nu,T}}{(2\pi)^{3/2}}  
 \at |\bk|^{Re\nu}+|\bk|^{-Re\nu} \ct
  | \widetilde{\ph}(\bk)|\:,
$$
  where the right hand side is integrable because $Re \nu <1/2$,   
   $ \widetilde{\ph} \in L^1(\bR^3;d\bk)\cap L^2(\bR^3;d\bk)$ and it vanishes  faster than any power for $|\bk|\to +\infty$. 
Lebesgue's dominate convergence theorem implies that the former limit in (\ref{2limits})
vanishes. The remaining limit has been computed in the proof of (a) in Theorem \ref{Nicola}.
The final result reads as follows: if $(\ell,\theta,\phi)$ are Bondi coordinates of $p\in \scrim$ and $\eta: \bS^2\to \bS^2$ is the inversion 
${\bf n} \mapsto -{\bf n}$ on the sphere,
 \beq
 \left(\gamma\Gamma \ph\right)(\ell,\theta,\phi)= i 
\frac{e^{-i\frac{\pi}{4}}}{(-\gamma)} \int_0^{+\infty} \sp\sp dk  \: \frac{e^{-i\ell k }}{\sqrt{2\pi}}\:\:
\sqrt{\frac{k}{2(-\gamma )}}\widetilde{\ph}\left( \frac{k}{(-\gamma)},\eta(\theta,\phi)\right) +  c.c.\:.\label{final2} 
\eeq
  From this point on the proof carries on up to the conclusions exactly as in the proof of (a) in 
Theorem \ref{Nicola}, since (\ref{ImsigmaM}) holds also in our generalised case, as (\ref{ImsigmaM2}) shows. 
$\Box$

\end{document}